\documentclass[fleqn,usenatbib]{mnras}

\usepackage{newtxtext,newtxmath}

\usepackage[T1]{fontenc}

\DeclareRobustCommand{\VAN}[3]{#2}
\let\VANthebibliography\thebibliography
\def\thebibliography{\DeclareRobustCommand{\VAN}[3]{##3}\VANthebibliography}

\usepackage{graphicx}
\usepackage{amsmath}
\usepackage{xcolor}
\usepackage{xspace}

\newcommand{\ergs}{\,erg\,s$^{-1}$\xspace}
\newcommand{\flux}{\,erg\,s$^{-1}$\,cm$^{-2}$\xspace}

\hypersetup{
    colorlinks=true,
    linkcolor=blue,
    filecolor=blue,      
    urlcolor=blue,
    citecolor=blue,
}

\title[The accretion flow and jet of M87]{The accretion flow and jet of M87: broadband modelling across spatial scales}

\author[E. Guti\'errez et al.]{
E. M. Guti\'errez,$^{1,2,3}$\thanks{E-mail: eduardo.gutierrez@aei.mpg.de}
S. del Palacio,$^{4}$
T. T. Mazzei$^{5,6}$
\\
$^{1}$Max-Planck-Institut für Gravitationsphysik (Albert-Einstein-Institut), Am Mühlenberg 1, D-14476 Potsdam-Golm, Germany\\
$^{2}$Institute for Gravitation and the Cosmos, The Pennsylvania State University, University Park, Pennsylvania, 16802, USA \\
$^{3}$Department of Physics, The Pennsylvania State University, University Park, Pennsylvania, 16802, USA\\
$^{4}$Department of Space, Earth and Environment, Chalmers University of Technology, SE-412 96 Gothenburg, Sweden\\
$^{5}$Facultad de Ciencias Astronómicas y Geofísicas, Universidad Nacional de La Plata, Paseo del Bosque s/n, 1900 La Plata, Buenos Aires, Argentina\\
$^{6}$Instituto de Astrofísica de La Plata (CCT La Plata, CONICET; UNLP), Paseo del Bosque s/n, 1900 La Plata, Buenos Aires, Argentina
}

\date{Accepted XXX. Received YYY; in original form ZZZ}

\pubyear{\the\year{}}

\begin{document}
\label{firstpage}
\pagerange{\pageref{firstpage}--\pageref{lastpage}}
\maketitle

\begin{abstract}
The nearby radio galaxy M87 is an ideal laboratory for studying accretion flows and relativistic jet emission.
The EHT collaboration conducted the most comprehensive multi-wavelength observational campaign of M87, in which radio instruments with resolutions ranging from tens of microarcseconds to about an arcsecond sampled different portions of the outflow.
We model these data using a radiatively inefficient accretion flow (RIAF) with winds and a new multi-zone jet model, initially magnetically dominated, in which we solve the transport equation for the non-thermal electrons as they are injected, advected, and cooled along the jet.
We then compare each radio flux with the jet emission within an effective aperture set by the angular scale of the corresponding observation and obtain the best fit for the jet parameters.
The comparison across spatial scales favours a scenario in which the RIAF, rather than the jet, produces the compact $230~{\rm GHz}$ emission, while the jet dominates the lower-frequency radio, optical/UV and X-ray emission.
The Jet+RIAF model connects the Bondi-scale gas supply to an accretion rate at the horizon of ${\approx}10^{-5}$ of the Eddington rate, consistent with EHT polarimetry, and the two jets carry in total ${\approx}13$ per cent of the accretion power, $10^{43}~{\rm erg~s}^{-1}$.
The framework presented in this work provides a simple way to disentangle the accretion-flow and jet contributions to the emission from nearby radio galaxies.
\end{abstract}


\section{Introduction}

Active galactic nuclei (AGN) with relativistic jets are among the most powerful and persistent particle accelerators in the Universe \citep{BlandfordKonigl1979, Blandford_etal2019, Krolik1999, Romero_etal2017, Matthews_etal2020NewAR..8901543M}.
Understanding how these jets are launched from the vicinity of a supermassive black hole (BH), how they are collimated and accelerated, and how they radiate across the electromagnetic spectrum remains an open problem in high-energy astrophysics.
Because of its proximity ($16.8\,$Mpc, \citealt{Mei_etal2007ApJ...655..144M, Blakeslee_etal2009ApJ...694..556B, Cantiello_etal2018ApJ...856..126C}) and the very massive central BH it hosts \citep[$6.5\times10^{9}\,M_\odot$,][]{Gebhardt_etal2011ApJ...729..119G, EHT_2019_VI_ApJ...875L...6E}, M87 is a cornerstone source for probing the physics of jet formation and the connection between accretion flows and outflows on horizon to kiloparsec scales \citep{Hada_etal2024A&ARv..32....5H}.  

Nearby radio galaxies are ideal laboratories for studying jets, since their inner regions can be revealed by very-long-baseline interferometry (VLBI) down to scales comparable to the size of the central BH, where the jet is launched, and the accretion flow radiates.
In M87, the BH is large enough on the sky for VLBI at millimetre wavelengths to resolve its immediate surroundings.
The Event Horizon Telescope (EHT) collaboration published the first image of the shadow of a BH, the one at the centre of M87, obtained from VLBI observations at 230\,GHz carried out in April 2017 \citep{eht2019}.
At the same time, the EHT Multi-wavelength Science Working Group coordinated the most complete quasi-simultaneous observing campaign on the M87 nucleus to date, from radio to very-high-energy (VHE) $\gamma$-rays \citep{EHTMWL2021}.
A second observational campaign in 2018 caught a VHE $\gamma$-ray flare \citep{EHTMWL2024}.
The radio data from these campaigns are especially useful because they provide angularly resolved information from event-horizon scales to hundreds of gravitational radii.

The 230\,GHz emission detected by the EHT is optically thin synchrotron radiation from the vicinity of the BH \citep{eht2019, eht2019_V}, but it is still debated whether it comes mainly from the outflow (the jet) or from the inflow (the accretion flow; e.g. \citealt{eht2019_V, Moscibrodzka_etal2016, Punsly2017, Emami_etal2021, Lu_etal2023, Yang_etal2024}).
The ring diameter, which is set by the BH's lensing, has remained stable across the 2017, 2018, and 2021 observations \citep{EHT_2024_2018image, EHT_2025_AA704}.
At 86\,GHz the ring is about 50 per cent larger than at 230\,GHz, and the jet base is wider than expected for a jet launched by the BH alone \citep{Lu_etal2023, Kim_etal2025}.
The 2021 EHT data also limit the 230\,GHz emission of the jet base to $\lesssim 60$\,mJy \citep{Saurabh_etal2026}.
General-relativistic magnetohydrodynamic (GRMHD) simulations with radiative transfer show that the answer may depend on the assumed electron heating and distribution \citep[e.g.][]{Moscibrodzka_etal2016, Davelaar_etal2019, Fromm_etal2022}.

Relativistic jets radiate synchrotron emission from relativistic electrons in strong magnetic fields; the emission is Doppler boosted and, at low frequencies, suppressed by synchrotron self-absorption \citep[SSA;][]{BlandfordKonigl1979, RybickiLightman1986}.
Jet emission is usually modelled either in the one-zone approximation, as a homogeneous relativistic blob \citep[e.g.][]{vanderLaan1966}, or with multizone models of an extended, inhomogeneous jet \citep[e.g.][]{BlandfordKonigl1979, PotterCotter2013}.
Both EHT multi-wavelength campaigns found that a single zone cannot account for the broadband emission of M87 \citep{EHTMWL2021, EHTMWL2024}.
In particular, one-zone models fail to reproduce the radio spectrum, in which the SSA opacity and hence the size of the emitting region change with frequency.

In a low-luminosity AGN, such as that in M87, the central BH is thought to be fed through a hot, radiatively inefficient accretion flow (RIAF, \citealt{NarayanYi1994, NarayanYi1995, NarayanYi1995b, YuanNarayan2014}) that starts to build up at distances of $r \lesssim R_{\rm B}$, where $R_{\rm B}\sim 7\times 10^5~ R_{\rm g}$ is the Bondi radius \citep{Bondi1952} and $R_{\rm g}=GM/c^2$ is the gravitational radius of the BH of mass $M$ (we also use the Schwarzschild radius, $R_{\rm S}=2R_{\rm g}$), and extends down to the event horizon.
The very hot electrons in this plasma may reach relativistic temperatures of up to $T_{\rm e}\gtrsim 10^{10}~{\rm K}$ and radiate through synchrotron radiation, bremsstrahlung radiation, and inverse Compton scattering.

The mass accretion rate reaching the BH is often substantially lower than that inferred at larger radii, owing to mass loss through winds driven within the accretion flow \citep{NarayanYi1995, Stone_etal1999, Yuan_etal2012}.
In M87 this is directly supported by observations.
The Bondi-scale accretion rate inferred from X-ray observations \citep{DiMatteo_etal2003ApJ...582..133D, Russell_etal2015MNRAS.451..588R} is two to three orders of magnitude larger than the accretion rate near the horizon allowed by EHT polarimetry \citep{eht2021_VIII, EHT_2023_IX}, and Faraday rotation measures indicate a density profile $\rho\propto r^{-1}$ at $5\times10^{3}$--$2\times10^{5}\,R_{\rm S}$, as expected from a flow with winds \citep{Park_etal2019ApJ871}.
A portion of the accreted energy is channelled into bipolar jets, which may be powered either by the rotational energy of the BH spacetime \citep{BlandfordZnajek1977} or by the rotating accretion disc \citep{BlandfordPayne1982}.
These jets can extend to scales of order Mpc and, along their length, undergo continuous energy exchange among magnetic, bulk kinetic, and internal components, ultimately releasing a fraction of this energy as radiation across a broad range of spatial scales and frequencies \citep[see][]{Blandford_etal2019}.

The spectral energy distribution (SED) of M87 has been modelled both with semi-analytic descriptions of the accretion flow and the jet \citep[e.g.][]{Feng_etal2016, FengWu2017, Lucchini_etal2019, Boughelilba_etal2022, Kuze_etal2024, Hirotani_etal2025} and with GRMHD simulations and radiative transfer \citep[e.g.][]{Moscibrodzka_etal2016, Ryan_etal2018, Chael_etal2019, Davelaar_etal2019, Fromm_etal2022, Niedzwiecki_etal2024}.
Several of these works used spatially-resolved constraints, such as the shape of the jet or the flux density, size and position of the radio core \citep{Moscibrodzka_etal2016, Chael_etal2019, Davelaar_etal2019, Lucchini_etal2019, Fromm_etal2022, Hirotani_etal2025}, and some noted that the fluxes of the SED are measured with very different angular resolutions \citep{Ryan_etal2018, Niedzwiecki_etal2024}.
However, none of them consistently compared the model output with the angular scales probed by the observations.

In this work, we develop a global inflow--outflow model of M87 that connects the accretion flow, from the Bondi radius down to the BH, with a jet that extends from its base to kiloparsec scales.
The accretion flow is an RIAF that loses mass through winds, anchored by the accretion rates inferred at the Bondi radius and near the horizon.
The jet is initially magnetically dominated and is gradually collimated and accelerated; along it, non-thermal electrons are continuously injected and transported as they cool via synchrotron radiation and adiabatic expansion.

Because the model predicts where along the jet the emission at each frequency is produced, it can be compared with each measurement on the spatial scale that the instrument actually probes.
For each instrument, we integrate the jet emission that falls within its effective aperture, from the compact region sampled by the EHT to the entire jet, and compare it with the 2017 EHT multi-wavelength data.
We use this to test two scenarios: one in which the jet produces all the emission (Jet-only), and one in which the RIAF is responsible for the compact 230\,GHz emission seen by the EHT and part of the optical and X-ray emission (Jet+RIAF).
We fit the jet parameters of both scenarios to the data with a Markov chain Monte Carlo (MCMC) method.

The paper is organised as follows.
In Section~\ref{sec:model}, we describe the RIAF, the jet, the transport and radiation of the non-thermal electrons, and the observational constraints that guide our modelling strategy.
In Section~\ref{sec:results}, we present the SEDs predicted by our model applied to M87 and compare them with the data on the corresponding spatial scales, discuss the physical conditions they imply in the jet and the accretion flow, and complement this with synthetic total intensity and spectral index maps.
We discuss our results and compare them with previous work in Section~\ref{sec:discussion}, and summarise our conclusions in Section~\ref{sec:conclusions}.

\section{Model}\label{sec:model}

\begin{figure*}
    \centering
    \includegraphics[width=\textwidth]{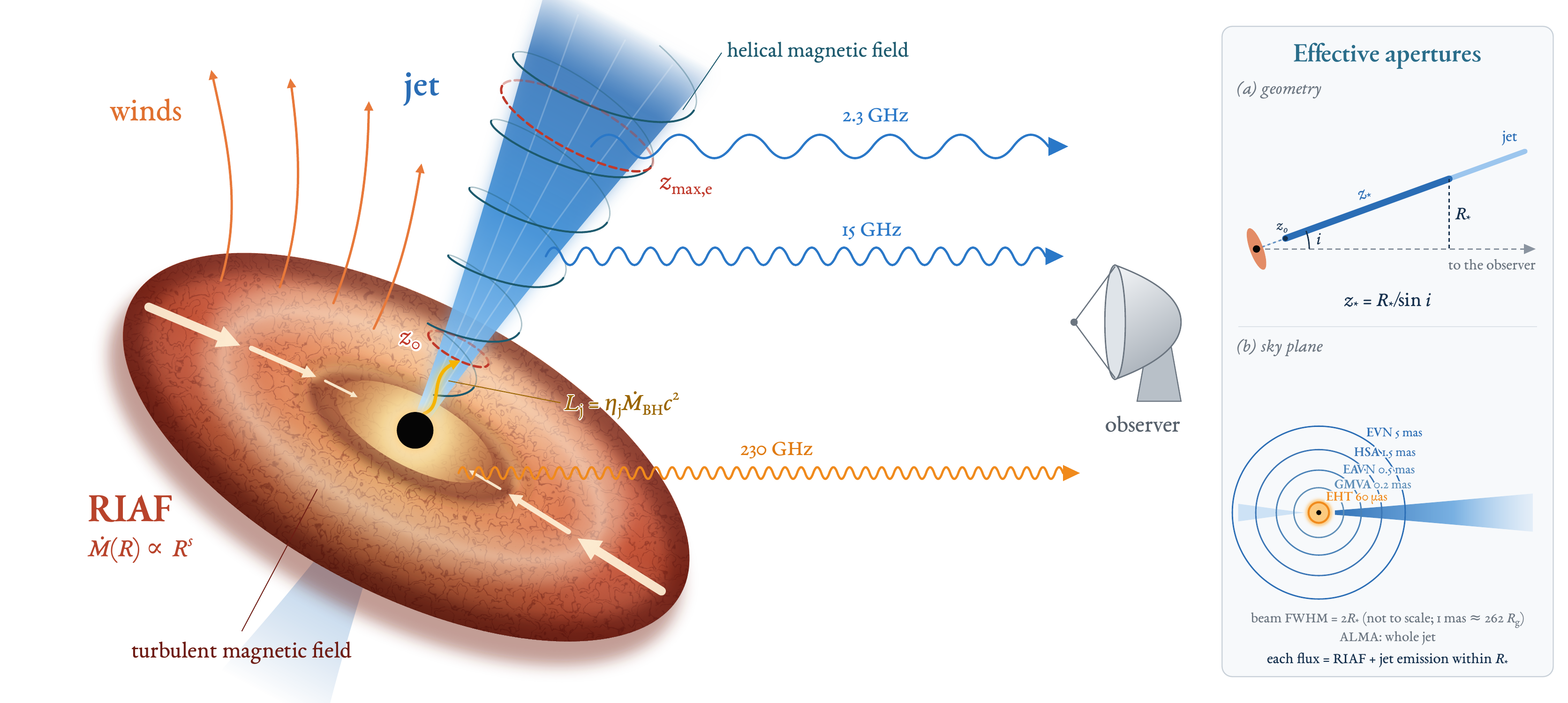}
    \caption{Schematic of the model.
    A geometrically thick RIAF with a turbulent magnetic field loses mass in winds, so that its accretion rate decreases inwards, $\dot{M}(R)\propto R^{s}$.
    A fraction $\eta_{\rm j}$ of the accretion power at the horizon, $L_{\rm j}=\eta_{\rm j}\dot{M}_{\rm BH}c^2$, powers each of the jet and the counter-jet, which are initially magnetically dominated.
    Non-thermal electrons are injected continuously between $z_0$ and $z_{\rm max,e}$ (dashed red rings) and cool while they are advected along the jet.
    The inner RIAF produces the compact $230$~GHz emission, whereas lower frequencies escape from the jet at increasing heights, as illustrated by the $15$ and $2.3$~GHz waves.
    Right: we approximate the region sampled by each observation with an effective aperture of radius $R_*$ (half the beam FWHM) on the plane of the sky (panel b, labelled with the beam FWHM).
    We include the full RIAF spectrum and the jet emission projected within $R_*$, approximately that up to the height $z_*=R_*/\sin i$ (panel a).}
    \label{fig:model}
\end{figure*}

In our model, the supermassive BH is fed by an RIAF that forms at large distances from it.
Outflows from the disc remove part of the inflowing gas, so that only a fraction of it reaches the BH.
The BH channels a fraction of the accreted energy into bipolar jets along its rotation axis.
The jets are magnetically dominated at their base; along them, magnetic energy is converted into bulk kinetic energy and into non-thermal particles, which are advected by the flow and cool through adiabatic work and synchrotron radiation, emitting from radio to X-rays.
Fig.~\ref{fig:model} illustrates the main components of the model and how the emission is compared with the observations. We describe each component below. 

\subsection{Accretion Flow} \label{subsec:accretion_flow}

We model the accretion flow as described in \citet{Gutierrez_etal2021} (see also \citealt{Manmoto_etal1997, YuanNarayan2014}), though restricting the treatment to purely thermal particles.
It consists of a hot, optically thin, geometrically thick, and radiatively inefficient flow, in which protons and electrons are not in thermal equilibrium.
Thermal electrons are energised both through ion coupling and direct heating by the flow turbulence, and radiate via synchrotron radiation, bremsstrahlung, and inverse Compton scattering.

We obtain the hydrodynamical structure of the RIAF by solving the height-integrated, steady-state hydrodynamic equations via the shooting method (e.g. \citealt{Abramowicz_etal1988, Yuan_etal2000, Yuan_etal2003}).
To estimate the cooling function for the flow, we follow the treatment in \citet{Manmoto_etal1997}.
The flow is determined by the following set of parameters: the outer accretion rate $\dot{M}_{\rm out}$ at radius $R_{\rm out}$, the $\alpha$ viscosity parameter, the gas pressure to magnetic pressure ratio, $\beta\equiv p_{\rm gas}/p_B$, and the fraction of energy released by turbulence directly heating electrons, $\delta_{\rm e}$.
In addition, the mass loss driven by magneto-centrifugal winds is included in the model through the parametrisation of the accretion rate $\dot{M}(R)=\dot{M}_{\rm out}(R/R_{\rm out})^s$, where $R_{\rm out}$ is the outer radius of the flow, $\dot{M}_{\rm out}$ is the accretion rate at that radius, and $0<s<1$ measures the wind strength \citep{BlandfordBegelman1999}.
Hence, the accretion rate onto the BH $\dot{M}_{\rm BH}$ can be much lower than that at the outermost boundary of the flow.

The hydrodynamic solution gives the temperature of electrons and ions, $T_{\rm e,i}(R)$, the magnetic field strength, $B(R)$, rest-mass density, $\rho(R)$, radial infall velocity, $v_R(R)$, and scale height, $H(R)$, as a function of the radial cylindrical coordinate.
With these properties, we compute the SED using the thermal emissivities by synchrotron \citep{Mahadevan_etal1996} and bremsstrahlung radiation \citep{Svensson1982, StepneyGuilbert1983, NarayanYi1995}.
Following \citet{Manmoto_etal1997}, we solve the radiative transfer in the vertical direction under the two-stream approximation \citep{RybickiLightman1986}.
To calculate the Comptonisation, we follow the approach by \citet{Narayan_etal1997b}, which iteratively couples the emission and scattering across different regions of the flow.
For more details, we refer the reader to \citet{Gutierrez_etal2021}.

\subsection{Jet} \label{subsection:jet}

Our jet model is partially based on the microquasar jet models by \citet{Zdziarskietal2014} and \citet{Molina_etal2019}.
We treat the jet as a continuous, one-dimensional flow, with thermodynamic properties derived from conservation laws and characterised by analytical functions of the height $z$ along the jet.
Relativistic electrons are injected along the jet and are advected by the flow tied to the magnetic field.
As they propagate, the electrons cool through various processes, mainly synchrotron radiation and adiabatic losses from the work they do to expand the fluid.
At high magnetisations, a potential mechanism for accelerating electrons is magnetic reconnection \citep{Matthews_etal2020NewAR..8901543M}.

We parametrise the jet width as a function of height as
\begin{equation}
    R_{\rm j}(z)=R_{{\rm j},0} (z/z_0)^w,
\end{equation}
where $R_{{\rm j},0}$ and $z_0$ are the half-width and height of the jet at its launching point, and $0\le w \le 1$ parametrises the jet geometry; values of $w=0, 0.5, 1$ correspond to a cylindrical, parabolic, and conical jet, respectively.

Similarly, we parametrise the Lorentz factor of the jet $\Gamma_{\rm j}$ as a power law, with an acceleration starting at a height $z_{\rm acc}$:
\begin{equation}
    \Gamma_{\rm j}(z) = \left[ 1 + \left( z/z_{\rm acc{}} \right)\right]^g,
\end{equation}
where the parameter $g$ measures how fast the jet accelerates.

The relativistic continuity equation, $\nabla_\alpha \left( \rho' u^\alpha \right) = 0$, implies that the mass flux,
\begin{equation}
    \dot{M}_{\rm j} = \pi R_{\rm j}^2 \Gamma_{\rm j} \beta_{\rm j} c \rho',
    \label{eq:Mdot}
\end{equation}
is constant along $z$.
Here, $u^\alpha$ is the fluid's four-velocity, $\rho'$ its comoving rest-mass density, and $\beta_{\rm j}=(1-\Gamma_{\rm j}^{-2})^{1/2}$.
Similarly, the conservation of the energy-momentum tensor, $\nabla_\alpha T^{\alpha \beta} = 0$,
implies that the total power of the jet,
\begin{equation}
    L_{\rm j} = \int_{\Sigma_z} T^{0z} d\Sigma_z = \Gamma_{\rm j} \dot{M}_{\rm j} c^2 h'(1+\sigma' ),
    \label{eq:Lj}
\end{equation}
is also conserved along the jet (if we neglect the radiation emitted by the jet particles).
In Eq. \ref{eq:Lj}, $\Sigma_z$ is the cross-section of the jet at height $z$, and $\sigma'$ and $h'$ are the magnetisation and the specific enthalpy, defined as
\begin{equation}
    \sigma' = \frac{B'^2}{4\pi \rho' h'c^2}, \quad \quad h'= 1 + \frac{e_{\rm e}' + p_{\rm e }'}{\rho'c^2},
    \label{eq:sigma_enthalpy}
\end{equation}
where $B'$ is the magnetic field, $e_{\rm e}'$ is the internal energy density, and $p_{\rm e}'$ is the gas pressure, all defined in the fluid's rest frame.

    From Eqs. \ref{eq:Mdot} and \ref{eq:Lj}, we can define the energy flux per unit of rest-mass energy flux, $\mu$, as
\begin{equation}
    \mu = \frac{L_{\rm j}}{\dot{M}_{\rm j} c^2} = \Gamma_{\rm j}  h'(1 + \sigma'),
    \label{eq:mu}
\end{equation}
which is also constant along the jet.

    Eq.~\ref{eq:mu} implies that a jet that is initially magnetically dominated, $\sigma'_0:=\sigma'(z_0) > 1$, can gain bulk kinetic energy and internal energy (accelerated particles) from the magnetic energy reservoir.
    Since $\Gamma_{\rm j}(z)$ is prescribed, Eq.~\ref{eq:mu} can only be satisfied with $\sigma'\geq0$ while $\Gamma_{\rm j}(z)h'(z)\le\mu$. Where this condition fails, the jet has run out of magnetic energy, and we set $\sigma'\simeq0$ (i.e., $B'\simeq0$).
    In the models of Sect.~\ref{sec:results} the condition holds along the whole jet.

    We assume that $h'(z_0)=1$, i.e., there are initially no relativistic electrons.
    We calculate the magnetic field strength at a height $z$ by combining Eqs. \ref{eq:sigma_enthalpy} and \ref{eq:mu}:
\begin{equation}
    B'(z) = \frac{2}{\Gamma_{\rm j}(z)R_{\rm j}(z)} \sqrt{\frac{L_{\rm j}}{\beta_{\rm j}(z ) c} \left( \frac{\sigma'(z)}{1+\sigma'(z)} \right) }.
\end{equation}

    Energetically, we can express the power of the jet as carried by three main components: cold matter, relativistic electrons, and magnetic fields:
\begin{equation}
    L_{\rm j} = L_{\rm p} + L_{\rm e} + L_B,
\end{equation}
where $L_{\rm p}$ is the rest-mass density (cold matter) power,
\begin{equation}
    L_{\rm p} = \pi R_{\rm j}^2 \Gamma_{\rm j}^2 \beta_{\rm j} c \rho' c^2 = \Gamma_{\rm j} \dot{M}_{\rm j}c^2,
    \label{eq:Lp}
\end{equation}
$L_{\rm e}$ is the power carried by the relativistic electrons,
\begin{equation}
   L_{\rm e} = \pi R_{\rm j}^2 \Gamma_{\rm j}^2 \beta_{\rm j} c \left( e_{\rm e}' + p_{\rm e}' \right),
   \label{eq:Le}
\end{equation}
where $e_{\rm e}'= n_{\rm e}' m_{\rm e} c^2 \langle \gamma_{\rm e}'-1 \rangle $ and $p_{\rm e}' =k_{\rm e} e_{\rm e}'$ are the energy density and pressure of the electrons, respectively.
    For relativistic particles, $k_{\rm e}=1/3$.
    Finally, the power carried by the magnetic field is
\begin{equation}
    L_B = \pi R_{\rm j}^2 \Gamma_{\rm j}^2 \beta_{\rm j} c \left( e_B' + p_B' \right),
    \label{eq:L_B}
\end{equation}
where, similarly, $e_B'=B'^2/8\pi$ is the magnetic energy density, and $p_B'$ is the magnetic pressure.
    For a fully ordered magnetic field, $p_B'=e'_B$.

We have assumed that the matter that carries most of the rest-mass flux $\dot{M}_{\rm j}$ is cold, without specifying whether it consists of an electron--proton or an electron--pair plasma.
In the models of Sect.~\ref{sec:results}, $\dot{M}_{\rm j}=L_{\rm p}/(\Gamma_{\rm j}c^2)\approx10^{-6}\,M_\odot\,{\rm yr}^{-1}$, and the number flux of relativistic electrons through the jet, $\pi R_{\rm j}^2\Gamma_{\rm j}\beta_{\rm j}c\int n'_{\rm e}\,d\gamma'_{\rm e}$, exceeds the proton flux that this $\dot{M}_{\rm j}$ would carry in an electron--proton plasma, $\dot{M}_{\rm j}/m_{\rm p}$, by a factor of ${\approx}80$--$280$ along most of the jet in the Jet+RIAF model, and of up to ${\sim}10^3$ near the base of the Jet-only model.
Most of these electrons, however, have Lorentz factors close to the assumed $\gamma'_{\rm e,min}\simeq2$, which the data do not constrain: those with $\gamma'_{\rm e}<10$ (about 90 per cent of them) do not contribute to the observed emission, and those with $10<\gamma'_{\rm e}<30$ only affect the self-absorption.
For the assumed $\sigma'_0=100$, the model jets must therefore contain pairs, at least a few per proton and ${\sim}10^2$ if the electron distribution extends down to $\gamma'_{\rm e}\simeq2$; a pair-dominated jet was also favoured by \citet{Reynolds_etal1996} from the properties of the self-absorbed radio core of M87, and pairs can be produced close to the BH, for example in magnetospheric gaps or by photon--photon collisions \citep[e.g.][]{LevinsonRieger2011, Moscibrodzka_etal2011, RomeroGutierrez2020}.
An electron--proton jet would have to be less magnetised: for $\sigma'_0=10$, the electrons with $\gamma'_{\rm e}>30$ no longer outnumber the protons.
We discuss hot protons in Sect.~\ref{subsec:caveats}.

    It is common to adopt an ad hoc prescription for how the electron energy density depends on $z$ (e.g., assuming that the electrons are in equipartition with the magnetic field; \citealt{BlandfordKonigl1979, Kaiser2006}).
    However, this approach does not self-consistently account for the evolution of relativistic electrons along the jet when cooling is included.
    Instead, here we impose the normalisation on the power injected in relativistic particles.
    We assume that a fraction $\epsilon_{\rm e}$ of the total jet power is transferred from the magnetic field to non-thermal electrons, $L_{\rm inj}=\epsilon_{\rm e}L_{\rm j}$, mostly in the region $z_0 < z < z_{\rm max,e}$.
    We parametrise the dependence of the injected power on $z$ as
\begin{equation}
    \label{eq:dLdz}
    \frac{dL_{\rm e}(z)}{dz} = \frac{L_0}{z_0} \left( \frac{z}{z_0} \right)^{-(1+a_{\rm e})} \exp\left(-\frac{z}{z_{\rm max,e}}\right), \quad z_0\le z\le z_{\rm max},
\end{equation}
where $z_{\rm max}$ is the end of the computed jet.
A value of $a_{\rm e}=0$ implies that for $z \ll z_{\rm max,e}$ equal power is injected per unit of $\log z$ along the jet (e.g., \citealt{Zdziarskietal2014}), whereas the exponential factor smoothly suppresses the injection beyond $z_{\rm max,e}$.
The total power injected up to a height $z$ is then
\begin{equation}
    \label{eq:power_injected}
    L_{\rm e}(<z)= \int_{z_0}^z d\tilde{z} \frac{dL_{\rm e}(\tilde{z})}{d\tilde{z}}.
\end{equation}
    The normalisation constant $L_{\rm 0}$ is found by imposing that the total injected power is $L_{\rm e}(<z_{\rm max})=L_{\rm inj}$, so that the power injected into the electrons up to a height $z$ is
\begin{equation}
    L_{\rm e}(<z) = \epsilon_{\rm e}L_{\rm j}\,
    \frac{\int_{1}^{z/z_0} x^{-(1+a_{\rm e})}\, e^{-x z_0/z_{\rm max,e}}\, dx}{\int_{1}^{z_{\rm max}/z_0} x^{-(1+a_{\rm e})}\, e^{-x z_0/z_{\rm max,e}}\, dx}.
    \label{eq:Le_z}
\end{equation}
If the exponential factor in Eq.~\ref{eq:dLdz} is replaced by a sharp cut-off at $z_{\rm max,e}$ (no injection beyond it), Eq.~\ref{eq:Le_z} reduces to
\begin{equation}
    L_{\rm e}(<z) = \epsilon_{\rm e}L_{\rm j}
    \begin{cases}
        \dfrac{1-(z_0/z)^{a_{\rm e}}}{1-(z_0/z_{\rm max, e})^{a_{\rm e}}}, & {\rm if}~a_{\rm e}\neq 0, \\[2ex]
        \dfrac{\ln(z/z_0)}{\ln(z_{\rm max,e}/z_0)}, & {\rm if} ~a_{\rm e}=0,
    \end{cases}
    \label{eq:Le_z_sharp}
\end{equation}
for $z_0\le z\le z_{\rm max,e}$.

\begin{figure}
    \includegraphics[width=0.98\linewidth]{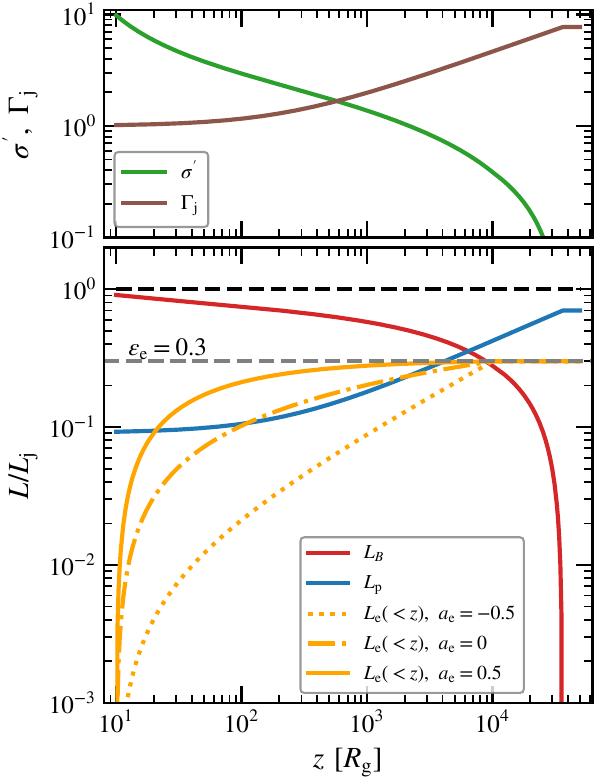}
    \caption{Evolution of the energy components along the jet for illustrative parameters: $\sigma'_0=10$, $g=0.4$, $z_{\rm acc}=220\,R_{\rm g}$, $z_0=10\,R_{\rm g}$, $\epsilon_{\rm e}=0.3$, and injection with a sharp cut at $z_{\rm max,e}=10^{4}\,R_{\rm g}$ (Eq.~\ref{eq:Le_z_sharp}) for $a_{\rm e}=-0.5$, $0$, and $0.5$. Top: Lorentz factor $\Gamma_{\rm j}$ and magnetisation $\sigma'$. Bottom: power in cold protons ($L_{\rm p}$), in the magnetic field ($L_B$), and injected into electrons up to $z$ ($L_{\rm e}(<z)$), normalised to $L_{\rm j}$.}
    \label{fig:jet_power}
\end{figure}

In Fig. \ref{fig:jet_power}, we show an example of how this set of equations dictates the evolution of the various components of the jet power.
The jet is magnetically dominated near its base (at $z \gtrsim z_0$), but the power in the magnetic field gradually decreases as it is converted into jet bulk kinetic power (Eq. \ref{eq:Lp}) and power in relativistic particles (Eq. \ref{eq:Le}).
We emphasise that the power injected into electrons up to a given height $z$ (given by Eq. \ref{eq:Le_z}) is not necessarily equal to the actual power carried by electrons at that height (given by Eq. \ref{eq:Le}), since part or most of the injected power was already radiated away below it. In other words, $L_{\rm e}(<z)$ is the sum of the power carried by the electrons at height $z$ and the power radiated between $z_0$ and $z$.

\subsubsection{Particle Energy Distribution} \label{subsec:SED_particulas}

At each height $z$, we assume that relativistic electrons are accelerated with a power-law spectrum with an exponential cutoff:
\begin{equation}
    q_{\rm e}'(\gamma', z) = q'_{{\rm e},0}(z) ~ \gamma_{\rm e}'^{-p} \exp \left[ -\gamma'_{\rm e}/\gamma'_{\rm e, cut}(z) \right],
\end{equation}
where $p$ is the spectral index of the distribution (taken constant along the jet), $\gamma'_{\rm e, cut}(z)$ is the cutoff energy of the electrons at height $z$, given by the balance between their cooling rate and the acceleration rate $\dot{\gamma}'_{\rm acc}=\eta_{\rm acc}eB'/(m_{\rm e}c)$, with $e$ the elementary charge and $\eta_{\rm acc}\le1$ the acceleration efficiency, and $q_{\rm e,0}'(z)$ is a normalisation factor.
Considering that $dL_{\rm e}/dz$ can be expressed as a function of the injection function as
\begin{multline}
    \frac{dL_{\rm e}(z)}{dz} = \Gamma_{\rm j} \pi R_{\rm j}^2 \int_{\gamma'_{\rm e,min}}^\infty q'_{\rm e}(\gamma'_{\rm e},z)\gamma'_{\rm e}m_{\rm e}c^2 d\gamma'_{\rm e} = \\
    \Gamma_{\rm j} \pi R_{\rm j}^2 m_{\rm e} c^2 q'_{{\rm e}, 0}(z) \int_{\gamma_{\rm e, min}'} ^{\infty} d\gamma_{\rm e}' ~ \gamma_{\rm e}'^{-p+1} \exp \left[ -\gamma'_{\rm e}/\gamma'_ {\rm e, cut}(z) \right],
\end{multline}
we can use Eq.~\ref{eq:dLdz} to obtain $q_{\rm e,0}'(z)$.
Here, we have taken $\gamma'_{\rm e,min}\simeq2$ as the minimum Lorentz factor of the injected electrons.

Assuming the system is in a steady state, the injected particles evolve in energy and height satisfying the transport equation:
\begin{multline}
    \frac{1}{\pi R_{\rm j}^2(z)} \frac{\partial}{\partial z} \left[ \pi R_{\rm j}^2(z) \Gamma_{\rm j}(z) \beta_{ \rm j}(z)~c~ n_{\rm e}'(\gamma'_{\rm e}, z) \right] +  \\  \frac{\partial}{\partial \gamma'_{\rm e}} \left[ \dot{\gamma}'_{\rm e}(\gamma'_{\rm e}, z)~ n_{\rm e}'(\gamma'_{\rm e},z) \right] = q_{\rm e}'(\gamma'_{\rm e} , z).
    \label{eq:transport_eq}
\end{multline}
Here, $n_{\rm e}'(\gamma'_{\rm e}, z)$ is the electron spectral particle density and $\dot{\gamma}'_{\rm e}$ is the rate of change of the comoving Lorentz factor of an electron owing to synchrotron and adiabatic losses.
Eq. \ref{eq:transport_eq} can be simplified by defining $\dot{\gamma}_{{\rm e},z}' := [\Gamma_{\rm j}(z)\beta_{\rm j}(z)c]^{-1 } \dot{\gamma}'_{\rm e}$, $dQ'_{\rm e}(\gamma'_{\rm e},z)/dz := \pi R_{\rm j}^2(z) q_{\rm e}'(\gamma'_{\rm e}, z)$, and $\tilde{N}'_{\rm e}(\gamma'_{\rm e}, z) := \pi R_{\rm j}^2(z) \Gamma_{\rm j}(z) \beta_{\rm j}(z) c n_{\rm e}'(\gamma'_{\rm e}, z)$, to obtain
\begin{equation}
    \frac{\partial \tilde{N}'_{\rm e}(\gamma'_{\rm e},z)}{\partial z} + \frac{\partial}{\partial \gamma'_{\rm e}} \left[ \dot{\gamma}_{{\rm e},z}'( \gamma'_{\rm e},z)~ \tilde{N}'_{\rm e}(\gamma'_{\rm e}, z) \right] = \frac{dQ'_{\rm e}(\gamma'_{\rm e}, z)}{dz}.
    \label{eq:transport_eq_mod}
\end{equation}
We solve this equation by discretising it with a fully implicit, first-order upwind finite-difference scheme with no diffusion term, i.e. the pure-cooling limit of the implicit schemes discussed by \citet{ParkPetrosyan1996} (see also \citealt{ChiabergeGhisellini1999}).
The scheme is unconditionally stable; we limit the step $\Delta z$ to 1 per cent of the cooling length of the electrons at the cooling break (the energy where the synchrotron cooling length equals the width of an output cell), and to the spacing of the output grid in $\log z$.

\subsubsection{Non-thermal Emission}\label{subsubsec:nonthermal}

Once we determine the particle energy distribution in each cell by solving Eq.~\ref{eq:transport_eq_mod}, we compute the electromagnetic emission from that cell.
In AGN jets, the relevant leptonic processes are synchrotron radiation and inverse Compton (IC) scattering of ambient photons, mainly the jet's own synchrotron photons (synchrotron self-Compton, SSC) and, near the base, those from the RIAF.
In the Thomson limit, both the IC-to-synchrotron cooling ratio and the SSC-to-synchrotron luminosity ratio equal the ratio of comoving photon to magnetic energy densities.
Because the jet is magnetically dominated, we estimate this ratio to be $\lesssim4\times10^{-3}$ for the synchrotron photons and $\lesssim2\times10^{-3}$ for the RIAF photons along the whole jet in the Jet+RIAF model, and $\lesssim 0.03$ for the synchrotron photons at the compact base of the Jet-only model.
IC cooling is therefore negligible (the electron distribution is shaped by synchrotron and adiabatic losses), and the SSC emission lies orders of magnitude below the observed X-ray and $\gamma$-ray fluxes.
We thus compute only the synchrotron emission, including synchrotron self-absorption.

We compute the synchrotron emissivity by integrating the single-particle synchrotron spectrum for the electron particle distribution:
\begin{equation}
    j'_{\nu'}(z)=\frac{1}{4\pi}\int_{\gamma'_{\rm e,min}}^\infty d\gamma'_{\rm e} ~n_{\rm e}'(\gamma_{\rm e}';z)~P_{\rm sy}(\nu';\gamma'_{\rm e},B'(z)),
\end{equation}
where we have assumed isotropic emission in the comoving frame, and we use the approximation given by \citet{Aharonian_etal2010PhRvD..82d3002A} for $P_{\rm sy}(\nu';\gamma'_{\rm e}, B'(z))$.
Relativistic electrons can also absorb emission from the jet via the SSA process.
We calculate the absorption coefficient for SSA in the jet's comoving frame using the formula given by \citet{RybickiLightman1986}.
Once we have the total emissivity, $j'_{\nu'}$ and the absorption coefficient $\alpha'_{\nu'}$ at a given height $z$, the observed energy flux on Earth is obtained by solving the radiative transfer equation.

A photon that leaves the jet at height $z$ towards an observer at an angle $i$ from the jet axis crosses a height interval $\Delta z\simeq R_{\rm j}(z)\cot i$. Assuming that the physical properties of the jet change little over this interval, i.e. $R_{\rm j}(z)\cot i\ll z$ (for a conical jet, $ i\gg \theta_{\rm j}$, with $\theta_{\rm j}$ the half-opening angle), the transport equation can be solved analytically for each cell, giving
\begin{equation}
    I_\nu(z,\psi)=\mathcal{D}_{\rm j}^3(z) S'_{\nu'}(z) \left[ 1-e^{-\tau'_{\nu'}(z,\psi)} \right],
\end{equation}
where $I_\nu(z, \psi)$ is the specific intensity, $\mathcal{D}_{\rm j}(z):=[\Gamma_{\rm j}(1-\beta_{\rm j} \cos i)]^{-1}$ is the Doppler factor, $S'_{\nu'}:= j'_{\nu'}/\alpha'_{\nu'}$ is the source function in the comoving frame, and
\begin{equation}
    \tau'_{\nu'}(z,\psi) = 
    \begin{cases}
        \dfrac{2R_{\rm j}(z)\alpha'_{\nu'}(z)}{\mathcal{D}_{\rm j}(z) \sin i}\sqrt{1-\psi^2}, & {\rm for}~|\psi| < 1, \\
        0, & {\rm otherwise},
    \end{cases}
\end{equation}
is the optical depth along the ray; here, $\psi:=x/R_{\rm j}$ and $x$ is the coordinate perpendicular to both the photon path and the jet axis \citep{Zdziarskietal2014}.
The observed energy flux is then
\begin{multline}
    F_\nu = \frac{\sin i}{d^2} \int_{z_{\rm min}}^{z_{\rm max}}dz~R_{\rm j}(z) \mathcal{D}_{\rm j}^3(z) S'_{\nu'}(z)\\
    \times  \int_{-1}^{1}d\psi\left\{1-\exp\left[-\tau'_{\nu'}(z,\psi)\right] \right\},
    \label{eq:obs_flux}
\end{multline}
Here, $d$ is the distance to the source, and the integral extends from the base of the jet, $z_{\rm min}=z_0$, to its end, $z_{\rm max}$; for the flux within an effective aperture, each cell is weighted by the fraction of its projection on the sky that falls within the aperture (Sect.~\ref{subsec:obs_constraints}). The observed and emitted frequencies are related simply by $\nu = \mathcal{D}_{\rm j}(z) \nu'$.
Close to the BH, we also include the gravitational redshift: we replace $\mathcal{D}_{\rm j}$ by $g_{\rm r}\mathcal{D}_{\rm j}$ in the equations above, where $g_{\rm r}=(1-2R_{\rm g}/r)^{1/2}$ and $r=(z^2+R_{\rm j}^2/2)^{1/2}$ is the root-mean-square distance to the BH of the jet cross-section at height $z$.
We neglect light bending, lensing and photon capture (Sect.~\ref{subsec:caveats}).
Equation \ref{eq:obs_flux} is valid for both optically thick and thin emission.
In particular, in the optically thin limit ($\tau'_{\nu'}\ll 1$), the expression simplifies to
\begin{equation}
    F_\nu \simeq \frac{\pi}{d^2} \int_{\rm z_{\rm min}}^{z_{\rm max}} dz  R_{\rm j}^2(z)
        \mathcal{D}_{\rm j}^2(z) j'_{\nu'}(z).
\end{equation}

Finally, we compute the emission of the counter-jet in the same way, assuming that it is identical to the jet but moving away from the observer, i.e. with $\mathcal{D}_{\rm cj}(z)=[\Gamma_{\rm j}(1+\beta_{\rm j}\cos i)]^{-1}$; all the fluxes shown below are the sum of both.

\subsection{Observational constraints and modelling strategy}\label{subsec:obs_constraints}

The parameters in our model are anchored by observational constraints on the mass inflow and jet energetics in M87.
\begin{itemize}
    \item[$\bullet$] X-ray observations of the hot gas at the Bondi radius provide an estimate of the outer mass supply \citep{DiMatteo_etal2003ApJ...582..133D, Russell_etal2015MNRAS.451..588R, Hada_etal2024A&ARv..32....5H}; we adopt an outer (Bondi) radius of $R_{\rm B}\simeq 7\times10^{5}~R_{\rm g}$, and an associated accretion rate $\dot{M}_{\rm B}\sim 1.4\times 10^{-3}~ \dot{M}_{\rm Edd}$, where $\dot{M}_{\rm Edd}=1.39\times 10^{18} (M/M_\odot)~{\rm g~s}^{-1}$.
    \item[$\bullet$] At much smaller radii, the EHT observations constrain the plasma properties near the horizon and imply a substantially lower accretion rate close to the BH, $\dot{M}_{\rm BH}\sim(2-14)\times 10^{-6}~\dot{M}_{\rm Edd}$ \citep{eht2021_VIII}.
    \item[$\bullet$] High-resolution radio imaging and kinematic studies supply information about the jet geometry $R_{\rm j}(z)$ \citep{Asada_etal2012ApJ...745L..28A, Hada_etal2013} and bulk speed  $\Gamma_{\rm j}(z)$ \citep{Mertens_etal2016A&A...595A..54M, Park_etal2019ApJ...887..147P}.
    The jet is approximately parabolic, for $z\gtrsim 100 R_{\rm g}$, following $R_{\rm j} \sim z^{0.56}$ \citep{Hada_etal2013}.
    The Lorentz factor of the jet $\Gamma_{\rm j}$ is observed to increase with the distance $z$ as a power law of index $g\sim0.16$ for $z>z_{\rm acc}\sim200R_{\rm g}$, at least up to the maximum heights we intend to treat in our model \citep{Park_etal2019ApJ...887..147P}.
    \item[$\bullet$] Estimates of the jet power span more than an order of magnitude. The energetics and age of the radio halo give a time-averaged power of $(6$--$10)\times10^{44}$\ergs over the last $\approx40$~Myr \citep{deGasperin_etal2012A&A...547A..56D}, whereas models of the compact emission of the jet give $\sim10^{43}$\ergs \citep[e.g.][]{Moscibrodzka_etal2016, Lucchini_etal2019, Davelaar_etal2019}, and GRMHD models of the EHT image require at least $\sim10^{42}$\ergs \citep{eht2019_V}.
\end{itemize}
These multi-scale measurements, therefore, set outer and inner anchors for the radial mass-flux as well as for the energetics available to the jet.
They also provide parameters for the geometric and dynamic evolution of the jet.
We tune the RIAF parameters by hand, as described below, and fit the jet parameters to the data (see the end of this section); both are listed in Table~\ref{tab:parameters}.

The two observational anchors for the accretion rate, $\dot{M}_{\rm B}$ (Bondi-scale X-rays) and $\dot{M}_{\rm BH}$ (EHT-scale constraints), bracket the allowed radial behaviour of $\dot{M}(R)$.
We adopt $\dot{M}_{\rm B}$ and $\dot{M}_{\rm BH}$ as the outer and inner mass-flux anchors, respectively.
Then, the wind index $s$ is constrained by
\begin{equation}
  s = \frac{\ln \left(\dot{M}_{\rm BH}/\dot{M}_{\rm B}\right)}{\ln \left(2R_{\rm g}/R_{\rm B}\right)} .
  \label{eq:s_from_mdot}
\end{equation}
In practice, we compute the RIAF structure only up to $R_{\rm out}=2\times10^{4}\,R_{\rm g}$, where we set $\dot{M}_{\rm out}$ (Table~\ref{tab:parameters}). With $s=0.4$ and $\dot{M}_{\rm out}=4.2\times10^{-4}\,\dot{M}_{\rm Edd}$, extrapolating $\dot{M}(R)$ to $R_{\rm B}$ gives $1.7\times10^{-3}\,\dot{M}_{\rm Edd}$, close to $\dot{M}_{\rm B}$, and the accretion rate at the inner boundary of the flow, $1.1\,R_{\rm S}$, is $\dot{M}_{\rm BH}=1.1\times10^{-5}\,\dot{M}_{\rm Edd}$, within the EHT range.
This two-anchor strategy (Bondi and EHT) provides a physically transparent way to connect the large-scale gas supply, the inner accretion state, and the multizone jet emission in a single, self-consistent framework.

With $\dot{M}(R)$ specified, we compute a self-consistent RIAF structure (density, temperatures, magnetic field) including mass loss, and compute the thermal emission by synchrotron and bremsstrahlung, as well as the Comptonised spectrum.
We follow the procedure implemented in \citet{Gutierrez_etal2021}, which closely follows \citet{Manmoto_etal1997}.

The available accretion power at the horizon, $L_{\rm acc}=\dot{M}_{\rm BH} c^{2}$,
provides a (soft) upper bound to the jet power through the so-called disc--jet connection, such that $L_{\rm j}=\eta_{\rm j}L_{\rm acc}$, where $\eta_{\rm j}<1$ is the jet efficiency (note, however, that in extreme circumstances, this parameter can be larger than $1$ since energy can be extracted from the BH rotation; \citealt{Tchekhovskoy2015}).
Then, observational bounds on $L_{\rm j}$ are used to constrain $\eta_{\rm j}$.
Finally, the jet is modelled as a multi-zone outflow with observationally motivated profiles $R_{\rm j}(z)$ and $\Gamma_{\rm j}(z)$ (quasi-parabolic).

We introduce a simple but effective method to account for the spatial scales probed by each radio instrument.
The radio fluxes of the EHT--MWL campaign are tagged by the angular scale of each observation \citep{EHTMWL2021}.
For each observation, we define an \emph{effective aperture}: an approximate integration region on the plane of the sky representing the angular scale associated with the reported flux.
For the radio arrays, the projected radius $R_*$ is half the FWHM of the circular beam with which the peak brightness was measured ($R_*=7.3\,R_{\rm g}$ for the EHT, about half the side of its integration region), converted to physical units ($1$~mas$\,\simeq262\,R_{\rm g}$ for our $M$ and $d$).
We compare each flux with the jet plus counter-jet emission that projects within the effective aperture. A cell of the jet at height $z$, with half-width $R_{\rm j}$, projects on to the sky as an ellipse centred at a distance $z\sin i$ from the BH, with semi-axes $R_{\rm j}$ across the jet and $R_{\rm j}\cos i$ along it; we count its emission with the fraction of the area of this ellipse that lies within a circle of radius $R_*$ around the BH, assuming a uniform surface brightness.
Where the jet is much narrower than the aperture, this amounts to integrating the jet emission up to the height $z_*=R_*/\sin i$; near the base, where the jet is wider than the smallest apertures, it excludes the part of each cell that projects outside them.
This sharp spatial cutoff approximates the observational selection, neglecting beam weighting and interferometric spatial filtering.
We add the full RIAF spectrum to every effective aperture, since the RIAF emission is compact.
We quantify the error of this approximation in Sect.~\ref{sec:results}.

In each scenario, we fit the eight jet parameters that shape the emission ($L_{\rm j}$, $\epsilon_{\rm e}$, $p$, $z_0$, $a_{\rm e}$, $z_{\rm max,e}$, $g$ and $\eta_{\rm acc}$) to the data, with the RIAF fixed.
The jet geometry ($w$ and the width at $z=140\,R_{\rm g}$), the onset of the bulk acceleration ($z_{\rm acc}=200\,R_{\rm g}$), $\sigma'_0$ and the remaining parameters are fixed at the values of Table~\ref{tab:parameters}.
We adopt the maximum-likelihood samples as the models of the two scenarios (Table~\ref{tab:parameters}); Sect.~\ref{sec:results} and Appendix~\ref{app:fits} describe the fit.

\section{Results}
\label{sec:results}

\begin{table*}
\centering
\caption{Parameters of the Jet+RIAF and Jet-only models.
For the eight fitted jet parameters, we give the prior range, the posterior median with the 16th and 84th percentiles and, in brackets, the maximum-likelihood (best-fitting) value, which defines the models shown in the figures (Sect.~\ref{subsec:obs_constraints} and Appendix~\ref{app:fits}); the quantities derived from them are given in the same way. Priors marked with $^\ast$ are uniform in the logarithm, and the others are uniform.
A single value applies to both models.
$L_{\rm j}$ and $\eta_{\rm j}$ refer to each of the two (jet and counter-jet) outflows.
The last column gives the reference for the parameters fixed from observations, or states whether a parameter is assumed (fixed by choice), tuned by hand to the SED, or derived from the other parameters.}
\label{tab:parameters}
\renewcommand{\arraystretch}{1.15}
\begin{tabular}{llllll}
\hline
Description & Parameter & Prior range & \textbf{Jet+RIAF} & \textbf{Jet-only} & Ref. \\
\hline
\multicolumn{6}{l}{\textbf{Black hole and global properties}} \\
Black hole mass & $M_{\rm BH}$ [$M_\odot$] & & \multicolumn{2}{c}{$6.5\times10^9$} & 1, 2 \\
Distance & $d$ [Mpc] & & \multicolumn{2}{c}{$16.8$} & 3, 4, 5 \\
Viewing angle & $i$ [deg] & & \multicolumn{2}{c}{$17$} & 6, 7 \\
\hline
\multicolumn{6}{l}{\textbf{RIAF}} \\
Accretion rate at $R_{\rm out}$ & $\dot{M}_{\rm out}$ [$\dot{M}_{\rm Edd}$] & & $4.2\times10^{-4}$ & --- & Tuned \\
Outer radius & $R_{\rm out}$ [$R_{\rm g}$] & & $2\times10^{4}$ & --- & Assumed \\
Mass-loss index & $s$ & & $0.4$ & --- & Tuned \\
Electron heating fraction & $\delta_{\rm e}$ & & $0.3$ & --- & Assumed \\
Viscosity parameter & $\alpha$ & & $0.1$ & --- & Assumed \\
Plasma beta, $p_{\rm gas}/p_B$ & $\beta$ & & $9$ & --- & Assumed \\
\hline
\multicolumn{6}{l}{\textbf{Jet: fixed parameters}} \\
Initial magnetisation & $\sigma'_0$ & & \multicolumn{2}{c}{$100$} & Assumed \\
Onset of bulk acceleration & $z_{\rm acc}$ [$R_{\rm g}$] & & \multicolumn{2}{c}{$200$} & 8 \\
Jet geometry parameter & $w$ & & \multicolumn{2}{c}{$0.56$} & 9 \\
Jet half-width at $z=140\,R_{\rm g}$ & $R_{\rm j}(140\,R_{\rm g})$ [$R_{\rm g}$] & & \multicolumn{2}{c}{$46$} & 9 \\
Length of the computed jet & $z_{\rm max}$ [$R_{\rm g}$] & & \multicolumn{2}{c}{$1.2\times10^{6}$} & Assumed \\
Electron energy range & $\gamma'_{\rm e,min}$, $\gamma'_{\rm e,max}$ & & \multicolumn{2}{c}{$2$, $10^{10}$\,$^{a}$} & Assumed \\
\hline
\multicolumn{6}{l}{\textbf{Jet: fitted parameters}} \\
Jet power (each jet) & $L_{\rm j}$ [$10^{42}$\ergs] & 1--100$^\ast$ & $5.2^{+4.7}_{-2.3}$ (5.6) & $8.8^{+8.6}_{-3.3}$ (6.1) & \\
Fraction of $L_{\rm j}$ injected into electrons & $\epsilon_{\rm e}$ & 0.001--0.3$^\ast$ & $0.084^{+0.093}_{-0.046}$ (0.090) & $0.16^{+0.09}_{-0.08}$ (0.26) & \\
Electron power-law index & $p$ & 1.9--2.6 & $2.15^{+0.03}_{-0.04}$ (2.19) & $2.13\pm0.02$ (2.15) & \\
Jet base height & $z_0$ [$R_{\rm g}$] & 2--50$^\ast$ & $13^{+14}_{-9}$ (19) & $2.27^{+0.42}_{-0.19}$ (2.01) & \\
Injection profile index & $a_{\rm e}$ & $-0.5$--2 & $-0.18^{+0.13}_{-0.15}$ ($-0.13$) & $0.71\pm0.09$ (0.72) & \\
Injection scale height (Eq.~\ref{eq:dLdz}) & $z_{\rm max,e}$ [$10^{4}\,R_{\rm g}$] & 0.1--10$^\ast$ & $1.9^{+3.3}_{-1.3}$ (6.0) & $2.3^{+4.1}_{-1.7}$ (2.4) & \\
Bulk acceleration parameter & $g$ & 0.10--0.20$^{b}$ & $0.16\pm0.03$ (0.20) & $0.16\pm0.03$ (0.19) & \\
Acceleration efficiency & $\eta_{\rm acc}$ & $10^{-4}$--0.1$^\ast$ & $0.018^{+0.036}_{-0.013}$ (0.069) & $0.046^{+0.034}_{-0.025}$ (0.065) & \\
\hline
\multicolumn{6}{l}{\textbf{Derived quantities}} \\
Accretion rate at ${\approx}R_{\rm S}$ & $\dot{M}_{\rm BH}$ [$\dot{M}_{\rm Edd}$] & & $1.1\times 10^{-5}$ & --- & Derived \\
Jet efficiency, $L_{\rm j}/(\dot{M}_{\rm BH}c^2)$, each jet & $\eta_{\rm j}$ [per cent] & & $5.9^{+5.3}_{-2.6}$ (6.3) & --- & Derived \\
Jet half-width at $z_0$ & $R_{{\rm j},0}$ [$R_{\rm g}$] & & $12^{+6}_{-5}$ (15) & $4.6^{+0.5}_{-0.2}$ (4.3) & Derived$^{c}$ \\
\hline
\multicolumn{6}{l}{\textbf{Fit}} \\
$\chi^2$ of the best fit ($N=45$ data points) & $\chi^2$ & & 27.0 & 44.2 & \\
Number of fitted parameters & $k$ & & 10$^{d}$ & 8 & \\
Reduced $\chi^2$ of the best fit & $\chi^2_{\rm red}=\chi^2/(N-k)$ & & 0.77 & 1.19 & \\
99 per cent upper limit on $z_0$ & $z_0$ [$R_{\rm g}$] & & --- & $<3.6$ & \\
EHT $230$~GHz flux, model/data & & & $1.06\pm0.01$ (1.05) & $0.72^{+0.11}_{-0.10}$ (0.78) & \\
ALMA $230$~GHz flux, model/data & & & $1.02^{+0.09}_{-0.07}$ (0.99) & $0.79\pm0.07$ (0.81) & \\
\hline
\end{tabular}

\medskip
\begin{minipage}{\textwidth}
\footnotesize
\textit{References.} (1) \citealt{Gebhardt_etal2011ApJ...729..119G}; (2) \citealt{EHT_2019_VI_ApJ...875L...6E}; (3) \citealt{Mei_etal2007ApJ...655..144M}; (4) \citealt{Blakeslee_etal2009ApJ...694..556B}; (5) \citealt{Cantiello_etal2018ApJ...856..126C}; (6) \citealt{Mertens_etal2016A&A...595A..54M}; (7) \citealt{Walker_etal2018}; (8) \citealt{Park_etal2019ApJ...887..147P}; (9) \citealt{Hada_etal2013}.

\textit{Notes.} $^{a}$Limits of the electron energy grid; the injected spectrum is cut off at $\gamma'_{\rm e,cut}(z)\approx(0.6$--$1.5)\times10^{7}$ at $z_0$, rising along the jet, set by the balance between acceleration and cooling; beyond $z\approx6.5\times10^{4}\,R_{\rm g}$, the cutoff reaches the upper limit of 200~TeV ($\gamma'_{\rm e}\approx4\times10^{8}$) set in the code, which affects only the emission above ${\sim}10^{21}$~Hz.
$^{b}$Range around the value $g\approx0.16$ measured by \citet{Park_etal2019ApJ...887..147P} (Sect.~\ref{subsec:obs_constraints}).
$^{c}$$R_{{\rm j},0}=R_{\rm j}(140\,R_{\rm g})\,(z_0/140\,R_{\rm g})^{w}$.
$^{d}$Including the two RIAF parameters that we tuned by hand, $\dot{M}_{\rm out}$ and $s$.
\end{minipage}
\end{table*}

\begin{figure*}
    \includegraphics[width=0.995\textwidth]{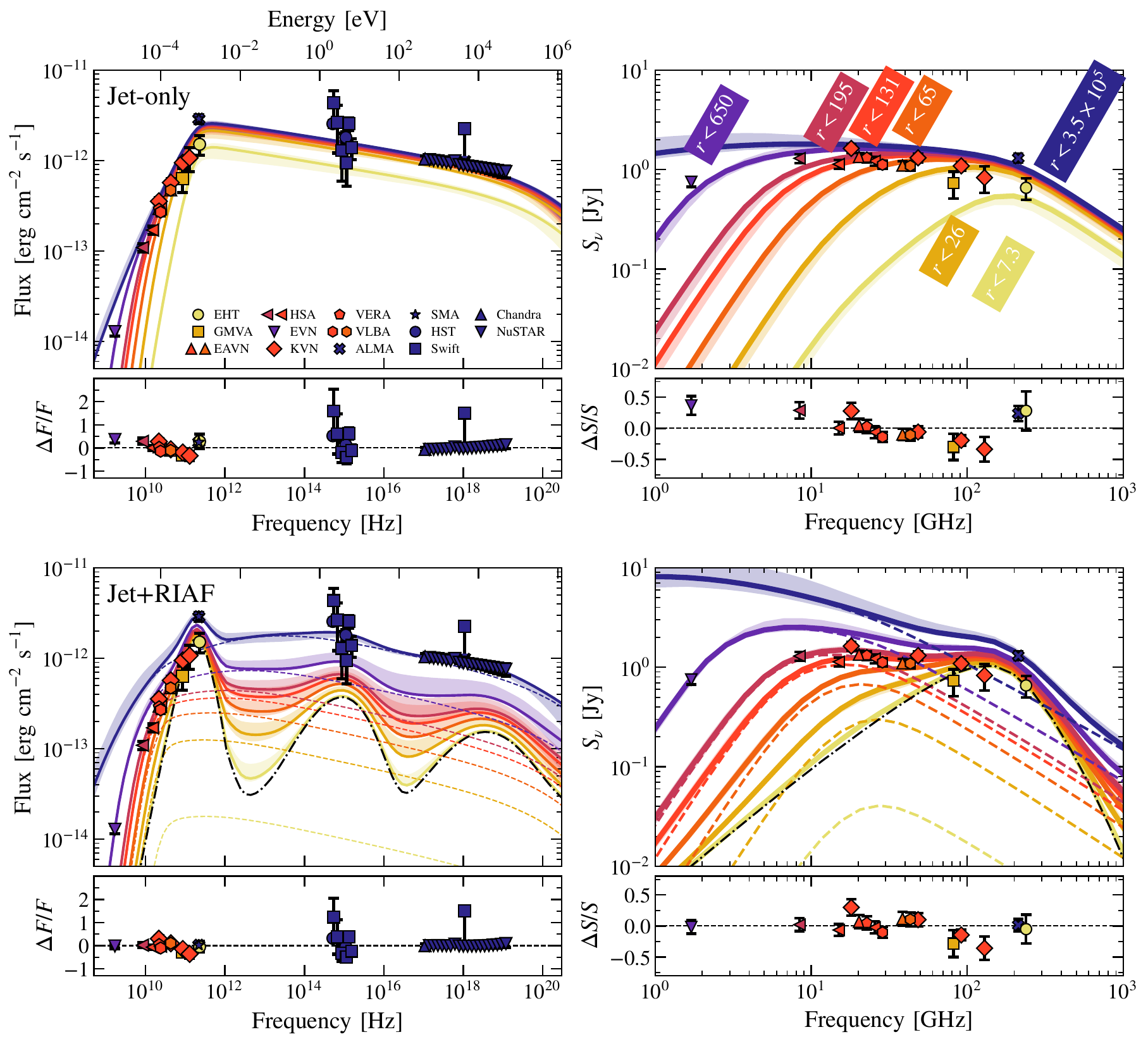}
    \caption{Spectral energy distribution of the best-fitting Jet-only model (top panels) and of the Jet+RIAF model (bottom panels) compared with the 2017 EHT--MWL data \citep{EHTMWL2021}.
    Left: broadband SED from radio to hard X-rays.
    Right: radio flux densities.
    The curves show the jet plus counter-jet emission within each of the seven effective apertures $R_*$ labelled in the top right panel, from $7.3$ to $3.5\times10^{5}\,R_{\rm g}$ (the last one includes the whole jet).
    In the bottom panels, dashed lines show the jet plus counter-jet emission for each effective aperture, the dash-dotted line the RIAF emission, and solid lines the total, in which the full RIAF spectrum is added to every effective aperture.
    The shaded bands show the 16th--84th percentile range of the total flux within each effective aperture over 200 samples of the posterior distribution (Appendix~\ref{app:fits}); the curves are the maximum-likelihood models.
    Each data point has the colour of the effective aperture used for its model comparison (Sect.~\ref{subsec:obs_constraints}); the SMA, optical/UV and X-ray data are compared with the whole jet and take the colour of the largest aperture.
    The symbols identify the instruments (instruments measured with two beams appear twice in the legend, once per effective-aperture colour).
    In the right panels, points at nearly the same frequency are displaced horizontally by up to $0.1$~dex for clarity.
    The lower panels show the residuals $(F_{\rm obs}-F_{\rm model})/F_{\rm model}$ for each point's effective aperture.}
    \label{fig:SEDs_M87}
\end{figure*}

We apply our model to M87 and consider two scenarios: one (Jet-only) in which the broadband emission is dominated by the jet and the accretion-flow emission is negligible, and another (Jet+RIAF) in which both components contribute.
In the Jet+RIAF scenario, the RIAF produces almost all the $230$~GHz flux measured by the EHT, because the jet contributes little within the EHT effective aperture; the two scenarios therefore correspond to the limiting cases of a negligible RIAF and of the brightest RIAF allowed by the EHT flux, and we do not explore intermediate cases with a fainter RIAF.
We compare each model with the fluxes at their corresponding spatial scales as described in Sect.~\ref{subsec:obs_constraints}.
Table~\ref{tab:parameters} lists the fitted parameters of both models: the posterior medians and the best-fitting (maximum-likelihood) values, which define the models shown below.

Fig.~\ref{fig:SEDs_M87} compares the Jet-only model (top panels) and the Jet+RIAF model (bottom panels) with the data, with each flux shown within its own effective aperture.
In particular, the models can reproduce the radio/mm fluxes observed at different angular scales.
We quantify the agreement with $\chi^2=\sum_i(F_{{\rm obs},i}-F_{{\rm mod},i})^2/\sigma_i^2$, where $\sigma_i^2=\sigma_{{\rm obs},i}^2+(0.1F_{{\rm obs},i})^2$ adds a systematic uncertainty of 10 per cent of the measured flux to the observational one; upper limits and $\gamma$-ray data are excluded, and no upper limit is exceeded by either model.
We fit the jet parameters with a Gaussian likelihood, $\ln\mathcal{L}=-\chi^2/2+{\rm const}$, and uniform priors over wide ranges (uniform in the logarithm for the scale parameters; Table~\ref{tab:parameters}), except for the bulk acceleration index, for which we use the narrow range $0.10\le g\le0.20$ around the value measured by \citet{Park_etal2019ApJ...887..147P}, and we sample the posterior distribution with the affine-invariant ensemble sampler \textsc{emcee} \citep{ForemanMackey_etal2013} through \textsc{bilby} \citep{Ashton_etal2019}.
The best-fitting Jet+RIAF model has a lower $\chi^2$ than the best-fitting Jet-only model, $\chi^2=27.0$ against $44.2$ for $N=45$ points, also after accounting for its two additional tuned parameters ($\dot{M}_{\rm out}$ and $s$): the reduced $\chi^2$, $\chi^2_{\rm red}\equiv\chi^2/(N-k)$, with $k$ the number of fitted parameters, is $0.77$ and $1.19$ for $k=10$ and $8$, respectively.
    
The optical and X-ray emission in the Jet-only model is reproduced by an optically thin synchrotron spectrum, $\nu F_\nu \sim \nu^{\alpha+1}$, with an index $\alpha=-\frac{p^*-1}{2}\approx -1$; here, $p^*=p+1\approx 3$ is the index of the electron particle distribution, $n'_{\rm e}(\gamma'_{\rm e})\sim\gamma_{\rm e}'^{-p^*}\sim \gamma_{\rm e}'^{-(p+1)}$, for synchrotron cooling-dominated electrons that are accelerated to a power-law with index $p\approx 2$.
Both models reproduce the {\it Chandra} and {\it NuSTAR} spectra and the {\it HST} flux at $1.1\times10^{15}$~Hz, but fall below the {\it HST} flux at $5.2\times10^{14}$~Hz, by 24 (Jet+RIAF) and 34 per cent (Jet-only), while the {\it Swift}/UVOT fluxes scatter around them (see Sect.~\ref{sec:discussion}).
The Jet+RIAF model has a similar jet synchrotron spectrum but benefits from the first and second Compton peaks of the RIAF spectrum, which contribute only a minor amount to the optical/UV and X-ray fluxes, slightly improving the agreement in the optical/UV.

The most important differences between the two models are seen in the radio data.
In both cases, the low-frequency emission is well-reproduced by the jet, which becomes optically thin at progressively lower frequencies at different heights.
However, the Jet-only model can approach the $230$~GHz flux within $7.3\,R_{\rm g}$ (EHT) only by placing the base of the jet at the lower limit of the prior, $z_0=2\,R_{\rm g}$, where a compact, self-absorbed component that peaks near 200~GHz produces it. Even so, it underestimates the $230$~GHz flux by about 20 per cent both within the EHT aperture and for the whole source (ALMA and SMA), and it fits the lower-frequency radio fluxes worse than the Jet+RIAF model.
In contrast, the Jet+RIAF model reproduces these data thanks to the synchrotron emission of the innermost regions of the RIAF, which peaks at around $230$~GHz.
In this scenario, the base of the jet ($z_0=19\,R_{\rm g}$) lies below the height $z_*=25\,R_{\rm g}$ that corresponds to the effective EHT aperture, but the jet emission within this aperture is weak, and the accretion flow provides 99 per cent of the EHT $230$~GHz flux.
Adding the whole RIAF spectrum to every effective aperture (Sect.~\ref{subsec:obs_constraints}) is a good approximation, since 91 per cent of the RIAF emission at $230$~GHz comes from within the effective EHT aperture deprojected onto the equatorial plane, and all of it at lower frequencies comes from within the larger effective apertures of the corresponding observations.

We note that our model is not intended to fit the $\gamma$-ray emission from the source: it includes only synchrotron emission (see Sect.~\ref{sec:discussion}), which drops sharply beyond the X-ray band.
In both models, the jet remains magnetically dominated along its whole length ($\sigma'\gtrsim6$ in the Jet+RIAF and $\sigma'\gtrsim2$ in the Jet-only model), so its inverse Compton emission is far too weak to account for the observed $\gamma$-rays (Sect.~\ref{subsec:gamma}).
In Appendix~\ref{app:scans}, we explore the dependence of the SED on each jet parameter (see Figs.~\ref{fig:scans_sed} and \ref{fig:scans_radio}), and in Appendix~\ref{app:fits} we give the posterior distributions of the parameters.
    
\subsection{Physical conditions in the jet and the accretion flow}\label{subsec:phys_cond}

\subsubsection{Magnetic field.} In the Jet+RIAF model, the comoving magnetic field at a projected distance of ${\sim}2$~mas is $B'\approx0.10$~G.
\citet{Ro_etal2023} inferred the comoving field at $2$--$10$~mas from the steepening of the $22$--$43$~GHz spectral index along the jet, which they attribute to synchrotron cooling, and found $B=(0.3$--$1.0~{\rm G})\,(z/2~{\rm mas})^{-0.73}$.
Our field declines with the same slope, since in both models $B'\propto(\Gamma_{\rm j}R_{\rm j})^{-1}$ (for $\sigma'\gg1$ in ours) with the same $R_{\rm j}(z)$ and $\Gamma_{\rm j}(z)$, but it is weaker by a factor of 3--10; the Jet-only model gives similar values.
In a jet with $\sigma'\gg 1$ and prescribed $R_{\rm j}(z)$ and $\Gamma_{\rm j}(z)$, $B'\propto L_{\rm j}^{1/2}$, and the fit constrains the jet power: over the posterior of the Jet+RIAF fit, $B'=0.11^{+0.04}_{-0.03}$~G at 2~mas, and only $0.4\%$ of the samples reach 0.3~G. The best model with $B'\ge0.3$~G among those computed during the sampling has $L_{\rm j}\approx4\times10^{43}$\ergs, $\epsilon_{\rm e}\approx6\times10^{-3}$ and $\chi^2_{\rm red}=1.06$, which is a better fit than that of the best Jet-only model ($\chi^2_{\rm red}=1.19$).

\subsubsection{Accretion rate and jet power.} In our Jet+RIAF model, the RIAF accretes $\dot{M}_{\rm BH}=1.1\times10^{-5}\,\dot{M}_{\rm Edd}$ ($1.6\times10^{-3}\,M_\odot\,{\rm yr}^{-1}$), within the range allowed by EHT polarimetry \citep{eht2021_VIII}, and radiates only a tiny fraction of the accretion power $\dot{M}_{\rm BH}c^2$.
Its accretion rate at ${\approx}21\,R_{\rm S}$ exceeds the upper limit that \citet{Kuo_etal2014} derived from the Faraday rotation measure of the $230$~GHz core, a limit derived assuming that the accretion flow is an external Faraday screen in front of the jet base, whereas in our model the $230$~GHz core is emitted by the accretion flow itself.
Each jet carries $L_{\rm j}=5.6\times10^{42}$~erg~s$^{-1}$ ($5.2^{+4.7}_{-2.3}\times10^{42}$~erg~s$^{-1}$ over the posterior), which corresponds to a jet efficiency $\eta_{\rm j}\approx6$ per cent per jet, and ${\approx}10$ per cent of this power is transferred to non-thermal electrons.
The jet accelerates to $\Gamma_{\rm j}\approx5.6$ and remains magnetically dominated along its whole length.

\subsection{Synthetic maps}

\begin{figure}
        \centering
        \includegraphics[width=0.99\linewidth]{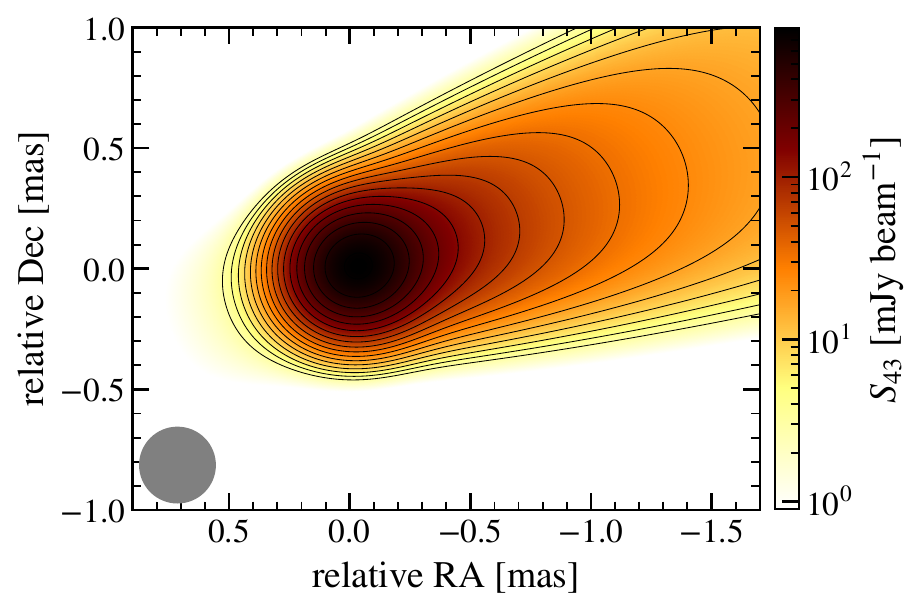}
        \caption{Synthetic $43$~GHz intensity map of the Jet+RIAF model (jet, counter-jet and RIAF). The emission is projected onto the plane of the sky, with the approaching jet at a position angle of $-72^\circ$ \citep{Walker_etal2018}, and convolved with a circular Gaussian beam of FWHM $0.31$~mas (grey disc).
        The black contours are at $3\times(1, \sqrt{2}, 2, 2\sqrt{2}, \dots)~{\rm mJy~beam}^{-1}$.
        The field, beam and contour levels are those of the VLBA $43$~GHz image of May 2017 in Fig.~14 of \citet{EHTMWL2021}, for direct comparison.}
        \label{fig:mapa_sintetico}
\end{figure}

\begin{figure*}
        \centering
        \includegraphics[width=0.85\textwidth]{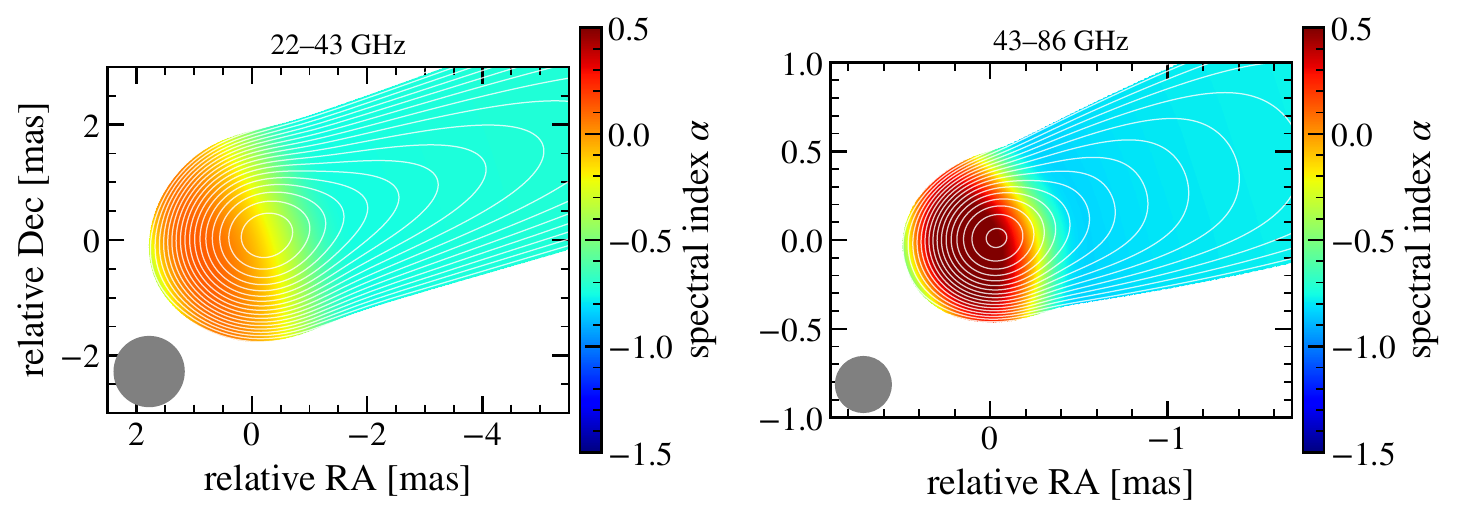}
        \caption{Synthetic spectral-index maps ($S_\nu\propto\nu^{\alpha}$) of the Jet+RIAF model between $22$ and $43$~GHz (left) and between $43$ and $86$~GHz (right).
        Both maps of each pair are convolved with the beam of the lower frequency (circular Gaussian of FWHM $1.2$ and $0.31$~mas, respectively; grey discs).
        Pixels where either map is below $3~{\rm mJy~beam}^{-1}$ are blank.
        The white contours show the intensity at the lower frequency, at $3\times(1, \sqrt{2}, 2, \dots)~{\rm mJy~beam}^{-1}$.
        The maps are shown on the sky, with the approaching jet at a position angle of $-72^\circ$, as in Fig.~\ref{fig:mapa_sintetico}.}
        \label{fig:map_spix}
\end{figure*}

Our model computes the emission at each height along the jet; thus, we can build synthetic maps for direct comparison with VLBI images.
We project the emission of the jet, the counter-jet and the RIAF (as a disc perpendicular to the jet, with the radial intensity profile of the RIAF model) on the plane of the sky, including synchrotron self-absorption along each line of sight through the jet, and convolve the result with a circular Gaussian beam similar to that of the observations we compare with.
Since the jet is one-dimensional, we assume that the non-thermal electrons and the magnetic field are uniform across each section of the jet, so the transverse brightness profile of the maps is set only by the path length through the jet.
The maps are therefore brightest along the jet axis and do not show limb brightening, but this approximation is sufficient to show how the brightness is distributed along the jet.

    Figure~\ref{fig:mapa_sintetico} shows the synthetic intensity map of the Jet+RIAF model at $43$~GHz, convolved with a beam of FWHM $0.31$~mas and drawn with the same field and contour levels as the VLBA image of the 2017 campaign \citep[][their fig.~14]{EHTMWL2021}.
The RIAF contributes about a quarter of the total flux density of the map, and the peak of the map lies close to the position of the BH.
The synthetic map closely resembles the observed brightness distribution, with a bright core and a jet that widens and fades westward.
Since the jet width follows the observed profile, this comparison mainly tests how the brightness is distributed along the jet.
The main difference is the limb-brightened structure of the observed jet, which a one-dimensional model cannot reproduce.

    Figure~\ref{fig:map_spix} shows the corresponding spectral-index maps.
We define the spectral index $\alpha$ by $S_\nu\propto\nu^{\alpha}$, i.e. $\alpha=\log(S_{\nu_2}/S_{\nu_1})/\log(\nu_2/\nu_1)$ between two frequencies $\nu_1$ and $\nu_2$, and compute it pixel by pixel from the two maps, both convolved with the beam of the lower frequency, as done for the observed maps.
For a homogeneous source of power-law electrons, $\alpha=5/2$ where the emission is optically thick (self-absorbed) and $\alpha=-(p-1)/2\approx-0.5$ where it is optically thin; synchrotron cooling steepens the optically thin spectrum down to $\alpha=-p/2\approx-1$, and the thermal electrons of the RIAF give $\alpha=2$ where they are optically thick.
Regions of different opacity within a beam give intermediate values.
Close to the BH, where the self-absorbed emission of the RIAF dominates, the spectrum is inverted between $43$ and $86$~GHz.
Between $22$ and $43$~GHz, it is nearly flat because the $1.2$~mas beam does not resolve the self-absorbed core and blends it with the optically thin emission of the jet; with the $0.31$~mas beam, the same pair of maps gives $\alpha\approx0.4$ at the core and up to $0.5$ next to it.
The observed maps show the same trend, a flat-to-inverted core and a steeper spectrum downstream \citep[][their fig.~15]{EHTMWL2021}.

\section{Discussion}
\label{sec:discussion}

\subsection{The origin of the 230~GHz emission}

Our fits favour the Jet+RIAF scenario over the Jet-only scenario.
The first one reproduces the compact $230$~GHz emission as coming almost entirely from the accretion flow, as well as the whole-source $230$~GHz flux, with the base of the jet anywhere between 2 and $50\,R_{\rm g}$ (Appendix~\ref{app:fits}).
A jet alone can approach the EHT flux only if it radiates from the immediate vicinity of the BH: the Jet-only fit puts the base of the jet at the lower limit of the prior, with $z_0<3.6\,R_{\rm g}$ at the 99 per cent level, and in the best-fitting model 83 per cent of the jet emission within the EHT aperture comes from $z<10\,R_{\rm g}$.
Even so, the Jet-only model falls ${\approx}20$ per cent short of the $230$~GHz fluxes on both scales and fits the data worse ($\Delta\chi^2\approx17$, with two parameters fewer).
The SED therefore favours the Jet+RIAF scenario, although on its own it does not exclude a jet origin of the compact $230$~GHz emission.
Observations at $86$~GHz support a large contribution of the accretion flow at millimetre wavelengths: the ring is about $50$ per cent larger and thicker than at $230$~GHz, and the jet base is wider than expected for a jet launched by the BH alone \citep{Lu_etal2023, Kim_etal2025}, although \citet{Hirotani_etal2025} attribute a ring of this size to the base of the jet.
In our Jet+RIAF model, the RIAF provides ${\approx}86$ per cent of the $86$~GHz flux within the GMVA effective aperture, and its emission is more extended at $86$ than at $230$~GHz, as the larger ring suggests.
At $230$~GHz, \citet{Saurabh_etal2026} found a jet-base component ${\approx}0.3$~mas from the BH with $\lesssim60$~mJy; in that region our jets emit ${\approx}34$ (Jet+RIAF) and ${\approx}26$~mJy (Jet-only), consistent with this limit.

As \citet{EHTMWL2024} concluded for the 2018 campaign, the jet dominates the SED as a whole: in our Jet+RIAF model, the RIAF provides ${\approx}40$ per cent of the total flux at $86$~GHz and ${\approx}55$ per cent at $230$~GHz, but only ${\approx}20$ per cent of the optical/UV flux and ${\approx}15$ per cent of the 2--10~keV flux.

\subsection{Accretion rate and winds}\label{subsec:mdot_winds}

Since the RIAF provides almost the whole EHT flux, its accretion rate near the BH, $\dot{M}_{\rm BH}=1.1\times10^{-5}\,\dot{M}_{\rm Edd}$ for the adopted $\delta_{\rm e}$ and $\beta$, lies in the upper part of the range allowed by EHT polarimetry, $(2$--$14)\times10^{-6}\,\dot{M}_{\rm Edd}$ \citep{eht2021_VIII}.
With $s=0.4$, less than $1$ per cent of the gas supplied at the Bondi radius reaches the BH.
For a self-similar flow, this wind index corresponds to a density profile $\rho\propto r^{-3/2+s}=r^{-1.1}$, close to the $\rho\propto r^{-1}$ inferred from Faraday rotation within the Bondi radius \citep{Park_etal2019ApJ871}, which supports mass loss of this strength independently of the SED.

\subsection{Jet power, efficiency and magnetic field}\label{subsec:jetpower}

The jet power of our models, ${\approx}6\times10^{42}$\ergs per jet in both best-fitting models ($(3$--$17)\times10^{42}$\ergs over the posteriors), is similar to that of other models of the compact emission \citep[e.g.][]{Moscibrodzka_etal2016, Lucchini_etal2019, Davelaar_etal2019} and much lower than the power averaged over the ${\approx}40$~Myr age of the radio halo \citep{deGasperin_etal2012A&A...547A..56D}, which need not reflect the present, historically low state of the source \citep{EHTMWL2021}.
The two jets of the Jet+RIAF model carry ${\approx}13$ per cent of the accretion power at the horizon (7--22 per cent over the posterior).
In simulations of magnetically arrested discs (MADs), which the EHT polarimetric data favour \citep{eht2021_VIII, EHT_2023_IX}, this efficiency corresponds to a slowly spinning BH with a prograde disc ($a_*\approx0.3$) or to a retrograde disc (7--19 per cent for $|a_*|=0.3$--0.9), whereas a prograde disc around a rapidly spinning BH is far more efficient (${\approx}130$ per cent for $a_*=0.9$; \citealt{Narayan_etal2022}).
\citet{FengWu2017}, who modelled the jet power with a variant of the Blandford--Znajek mechanism around an advection-dominated accretion flow (ADAF), inferred instead a rapidly spinning BH: since the power extracted from the BH grows with the magnetic flux at the horizon, which is largest in a MAD, the spin inferred from a given jet power depends on the assumed accretion state.

\subsection{$\gamma$-ray emission}\label{subsec:gamma}

The SSC emission of our jets is negligible.
In the Thomson regime, its ratio to the synchrotron luminosity equals the ratio of the comoving energy densities of the radiation and of the magnetic field, $e'_{\rm rad}/e'_{B}$, which is $\lesssim4\times10^{-3}$ along the jet of the Jet+RIAF model and $\lesssim0.03$ at the base of the Jet-only model (Sect.~\ref{subsubsec:nonthermal}) because the electrons receive only a fraction $\epsilon_{\rm e}\approx0.1$--$0.3$ of a jet power carried mostly by the magnetic field.
The observed $\gamma$-ray flux, $\nu F_\nu\approx(2$--$5)\times10^{-12}$\flux \citep{EHTMWL2021}, is instead comparable to the peak of the synchrotron emission and exceeds the X-ray flux, so an SSC origin would require $e'_{\rm rad}\sim e'_{B}$, i.e.\ a plasma far from magnetic dominance.
This agrees with \citet{Lucchini_etal2019}, who concluded that the $\gamma$-rays do not originate in the magnetically dominated inner jet.
The variable $\gamma$-ray emission, which extends to the TeV range \citep{EHTMWL2024, HESS_2024, LHAASO_2024}, thus requires another region or process, such as particle acceleration by magnetic reconnection close to the BH \citep{Kuze_etal2024, Hakobyan_etal2025}, regions further downstream where the particles carry a larger fraction of the energy, or hadronic emission \citep[e.g.][]{Boughelilba_etal2022}, for example from proton--proton interactions in the jet \citep{Xue_etal2022, Xue_etal2025}.

\subsection{Comparison with other models}

Semi-analytic models of the SED of M87 differ mainly in the origin they assign to the millimetre emission.
In jet-only models, it comes from the base of the jet.
\citet{Lucchini_etal2019} fitted the core SED together with the jet shape using a multi-zone jet model, parametrised on GRMHD simulations, in which thermal electrons produce the sub-millimetre bump; our Jet-only model resembles theirs in magnetisation and origin of the X-rays, but without thermal electrons it approaches the $230$~GHz flux only with its base at $z_0\approx2\,R_{\rm g}$. \citet{Hirotani_etal2025} fitted the radio spectrum and the core shift with an analytic model of a stationary, axisymmetric jet, constrained by VLBI data and GRMHD simulations; the base of their jet, at ${\approx}7\,R_{\rm S}$, produces a ring-like structure at $230$~GHz, but they treated the optical-to-X-ray fluxes as upper limits, whereas our jet must also produce them.
In semi-analytic models that combine an accretion flow with a jet, the millimetre emission instead comes from the inner accretion flow, as in our Jet+RIAF model.
With an ADAF plus jet model, \citet{Feng_etal2016} found a wind index ($s\approx0.4$--$0.5$) and an accretion rate near the BH (${\approx}10^{-3}\,M_\odot\,{\rm yr}^{-1}$) similar to ours, with a jet of a few $10^{42}$\ergs, comparable to ours, that produces the low-frequency radio and X-ray emission.
\citet{Kuze_etal2024} used a two-zone model in which a MAD produces the millimetre and $\gamma$-ray emission and a one-zone jet the optical and X-ray emission; a single jet zone, however, cannot reproduce the radio spectrum.

Our model differs from these works in two main respects.
First, apart from the one-zone jet of \citet{Kuze_etal2024}, these jets are multi-zone but prescribe the electron distribution at each height, whereas we solve the transport equation of electrons that are injected continuously and cool while they are advected along the jet, so that the distribution at each height is self-consistently derived from the available power carried by the jet.
Second, they compare the model with a single SED, that of the $0.4''$ core \citep{Feng_etal2016, Lucchini_etal2019} or that of the campaign as a whole (\citealt{Kuze_etal2024}; \citealt{Hirotani_etal2025} also use the core shift), whereas we compare each flux with the emission within the effective aperture of the corresponding observation.

One-zone models of the campaign data need a compact, strongly magnetised zone ($B'\approx5$--$10$~G) for the compact emission and either a larger, particle-dominated zone \citep{EHTMWL2021, EHTMWL2024} or hadronic emission \citep{Boughelilba_etal2022} for the $\gamma$-rays; \citet{Boughelilba_etal2022} attributed the X-rays to the accretion flow.
Our jets have fields of this order at their base ($B'\approx4$ and $23$~G at $z_0$ in the Jet+RIAF and Jet-only models, respectively), but the field decreases along the jet, so that each frequency is emitted on a different scale. A homogeneous zone cannot reproduce the frequency-dependent size of the radio emission, as constrained by measurements at different angular scales.
In our model, the RIAF produces the compact millimetre emission and a single extended jet reproduces the radio fluxes within all the effective apertures and the X-ray spectrum, while falling somewhat below the optical/UV fluxes (Sect.~\ref{subsec:caveats}); only the $\gamma$-rays require a separate component (Sect.~\ref{subsec:gamma}).

GRMHD simulations describe the dynamics of the accretion flow and the jet base self-consistently and in general relativity, which our model prescribes or neglects \citep[e.g.][]{Fromm_etal2022, Chael_etal2025}.
They are, however, limited to a region close to the BH, whereas our jet extends to ${\sim}10^{6}\,R_{\rm g}$ and covers all the spatial scales sampled by the campaign.
In addition, they either prescribe the electron distribution in each cell, through the electron-to-ion temperature ratio or a non-thermal ($\kappa$) distribution \citep{Davelaar_etal2019, Fromm_etal2022}, or evolve only thermal electrons \citep{Chael_etal2025}, rather than following the transport and cooling of accelerated electrons along the jet.
\citet{Niedzwiecki_etal2024} found that a one-dimensional ADAF model cannot reproduce the whole SED of the core, whereas in their GRMHD model the flow within $4\,R_{\rm g}$ produces the millimetre-to-UV emission and a non-thermal tail of the electron distribution produces the X-rays.
In our model, the RIAF only needs to reproduce the millimetre emission, while it only marginally contributes to the optical/UV and X-ray emission.

\subsection{Caveats of our model}\label{subsec:caveats}

The jet is one-dimensional, so it has no transverse structure: it cannot reproduce the limb-brightened morphology or the helical threads seen in VLBI images \citep[e.g.][]{Nikonov_etal2023}.
We have also restricted our models to magnetically dominated jets with leptonic emission, in which the protons, if present, are cold. Depending on the acceleration mechanism, protons may gain as much energy as the electrons.
If a fraction $\epsilon_{\rm p}$ of the jet power is given to relativistic protons with the same injection profile as the electrons, and they do not cool, their enthalpy lowers the magnetisation of the jet.
For $\epsilon_{\rm p}\le0.1$, this lowers $B'$ along the jet by at most 8 per cent, which a jet power up to ${\approx}17$ per cent larger compensates, since $B'\propto L_{\rm j}^{1/2}$, and the jet remains magnetically dominated.
Hot protons are therefore degenerate with the jet power, and would not change our conclusions; they would, however, increase the jet power that the SED requires, and they could produce $\gamma$-rays through hadronic interactions \citep{Xue_etal2022, Xue_etal2025}.
Finally, we keep the parabolic shape out to the end of the computed jet, whereas the observed jet becomes conical upstream of HST-1 \citep{Asada_etal2012ApJ...745L..28A}; this only affects the fluxes that we compare with the whole jet.

For M87, the condition $R_{\rm j}\cot i\ll z$ holds only at $z\gtrsim10^4\,R_{\rm g}$; closer to the BH, the line of sight crosses a height range comparable to $z$.
We account for the extent of each section on the sky in the effective apertures (Sect.~\ref{subsec:obs_constraints}), but not for this crossing, which mainly affects the self-absorbed base of the jet: in the Jet+RIAF model, the jet is optically thin at all heights for frequencies above $22$~GHz and self-absorbed only at $z\lesssim700\,R_{\rm g}$ at $5$--$15$~GHz, and a ray-tracing calculation through the same jet (in flat space) would change its fluxes by a factor comparable to the adopted uncertainties.
The base of the Jet-only model is self-absorbed up to ${\approx}230$~GHz and much more affected:
light bending, lensing and photon capture, which we neglect, could change the emission from within a few $R_{\rm g}$ of the BH.

Both of our models fall below the HST flux of the core at $5.2\times10^{14}$~Hz by 24--34 per cent, while the \textit{Swift}/UVOT fluxes scatter around them.
The HST fluxes depend on the subtraction of the bright host galaxy, which dominates their uncertainty \citep{EHTMWL2021} and can leave residual starlight in the nuclear aperture \citep{Prieto_etal2016, Lucchini_etal2019, Roder_etal2025}.
Other possible causes include variability between observations of different bands or components missing from our models, such as non-thermal electrons in the RIAF \citep{Gutierrez_etal2020, Gutierrez_etal2021, Niedzwiecki_etal2024}.

\section{Conclusions}
\label{sec:conclusions}

We have modelled the broadband emission of M87 during the 2017 EHT multi-wavelength campaign with an RIAF with winds and a one-dimensional, multi-zone jet that is initially magnetically dominated.
Non-thermal electrons are injected continuously and transported with radiative and adiabatic cooling.
For each radio instrument, we integrated the jet emission within its effective aperture, from $7.3\,R_{\rm g}$ for the EHT to the whole jet.

We fitted the jet parameters of a Jet-only and of a Jet+RIAF scenario. Both broadly reproduce the radio-to-X-ray emission, but the data favour the Jet+RIAF scenario, which reproduces the $230$~GHz fluxes on both the EHT and the arcsecond scales with the base of the jet anywhere between 2 and $50\,R_{\rm g}$, and better describes the data across spatial scales.
In this scenario, the RIAF produces the compact 230~GHz emission seen by the EHT, while the jet dominates the radio emission at lower frequencies and the optical/UV and X-ray emission, and its emission near the base is consistent with the EHT limit on a jet-base component.
A RIAF with a wind index $s=0.4$ connects the Bondi-scale accretion rate to $\dot{M}_{\rm BH}=1.1\times10^{-5}\,\dot{M}_{\rm Edd}$, within the range allowed by EHT polarimetry, with a density profile close to that inferred from Faraday rotation.
Each jet carries $5.6\times10^{42}$\ergs, ${\approx}6$ per cent of the accretion power at the horizon.
For the assumed magnetisation, $\sigma'_0=100$, the number of radio-emitting electrons implies that the jets contain pairs, although a less magnetised electron--proton jet would fit the data similarly; hot protons would not change these results.
The fit also constrains the injection of non-thermal electrons along the jet: they receive about a tenth of the jet power with a spectral index $p=2.15^{+0.03}_{-0.04}$, and the injection profile index, $a_{\rm e}=-0.18^{+0.13}_{-0.15}$, is close to $a_{\rm e}=0$, for which the same power is injected per logarithmic interval of height.
Neither model produces the $\gamma$-rays, which must come from outside the magnetically dominated jet or from other processes.

We conclude that modelling the VLBI fluxes within their respective effective apertures provides a simple way to separate the accretion-flow and jet contributions to the emission from nearby radio galaxies.
The same model also yields synthetic intensity and spectral-index maps that can be compared directly with VLBI images; in future work, we will extend it to include the jet's transverse structure, which is needed to describe its limb-brightened morphology.

\section*{Acknowledgements}
We thank Luciano Combi, Federico L\'opez Armengol, and Florencia Vieyro for helpful discussions.
E. M. G. acknowledges the use of Claude models by Anthropic and ChatGPT models
by OpenAI to accelerate code development and manuscript preparation.
T. T. M. acknowledges financial support from the Consejo Interuniversitario Nacional (CIN) through the Scientific Vocations Encouragement Scholarship (EVC-CIN) and from a CONICET (Argentina's National Research Council) doctoral fellowship.

\section*{Data Availability}

The jet code ExtendedJet is publicly available at \url{https://github.com/santimda/ExtendedJet}, and the RIAF code at
\url{https://github.com/eduardomgutierrez/thermal-riaf}.
The model configurations, the observational data points, the scripts that produce the figures and numbers of the best-fitting models, and the script that runs the fits of Appendix~\ref{app:fits} are available at \url{https://github.com/eduardomgutierrez/m87-riaf-jet-repro}.
That repository pins both codes at the exact versions used here.

\bibliographystyle{mnras}
\bibliography{biblio}

\appendix

\section{Jet model parameters}
\label{app:scans}

\begin{figure*}
    \centering
    \includegraphics[width=0.995\textwidth]{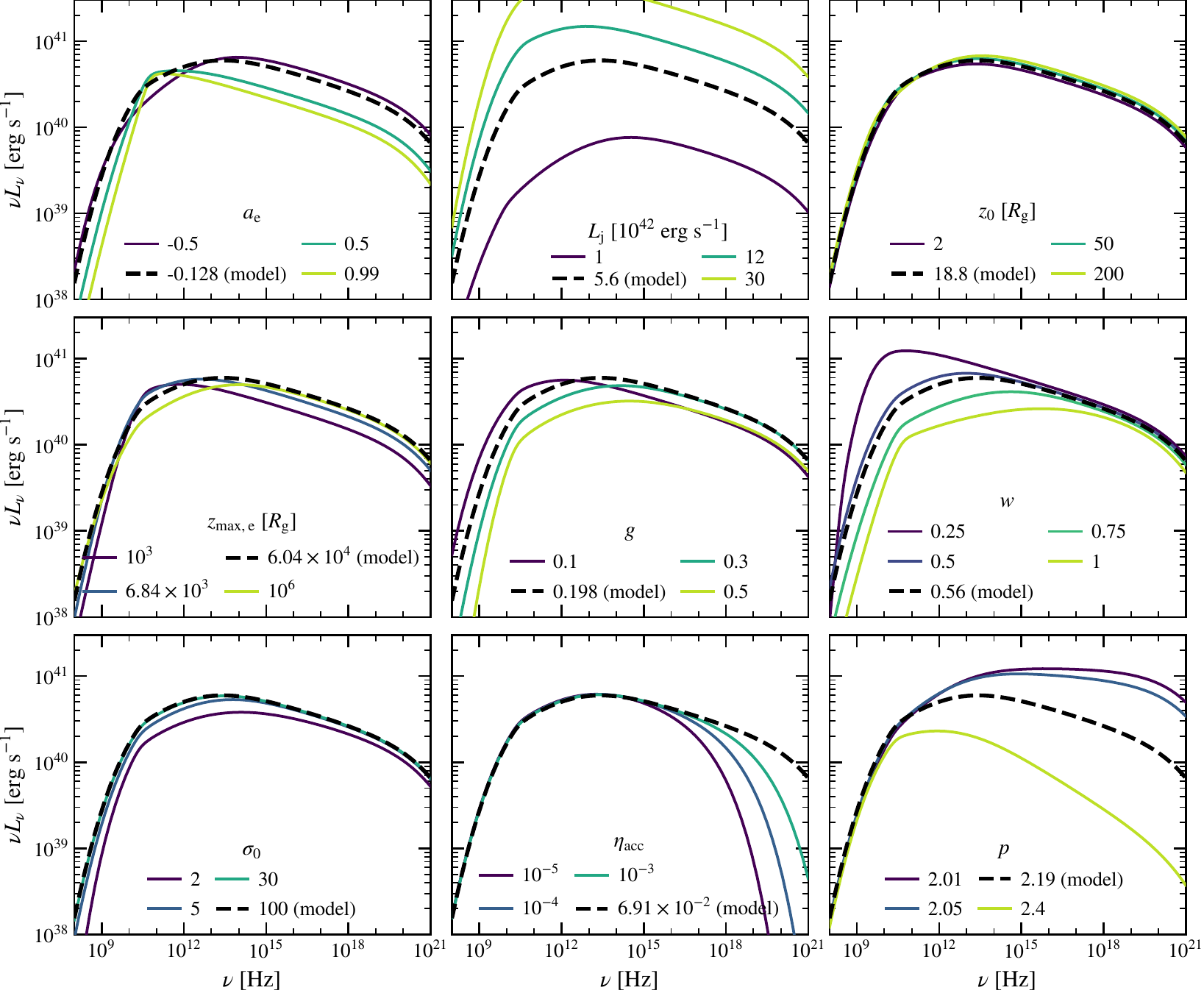}
    \caption{Jet parameter exploration: broadband SED ($\nu L_\nu$) of the jet plus counter-jet, integrated over the whole jet, when one parameter is varied, and all others are kept at the best-fitting Jet+RIAF values of Table~\ref{tab:parameters} (including the gravitational redshift; $\epsilon_{\rm e}=0.090$ is not varied). The black dashed curve is that model. The RIAF emission is not included. Panels, left to right and top to bottom: $a_{\rm e}$, $L_{\rm j}$, $z_0$, $z_{\rm max,e}$, $g$, $w$, $\sigma'_0$, $\eta_{\rm acc}$ and $p$.}
    \label{fig:scans_sed}
\end{figure*}

\begin{figure*}
    \centering
    \includegraphics[width=0.995\textwidth]{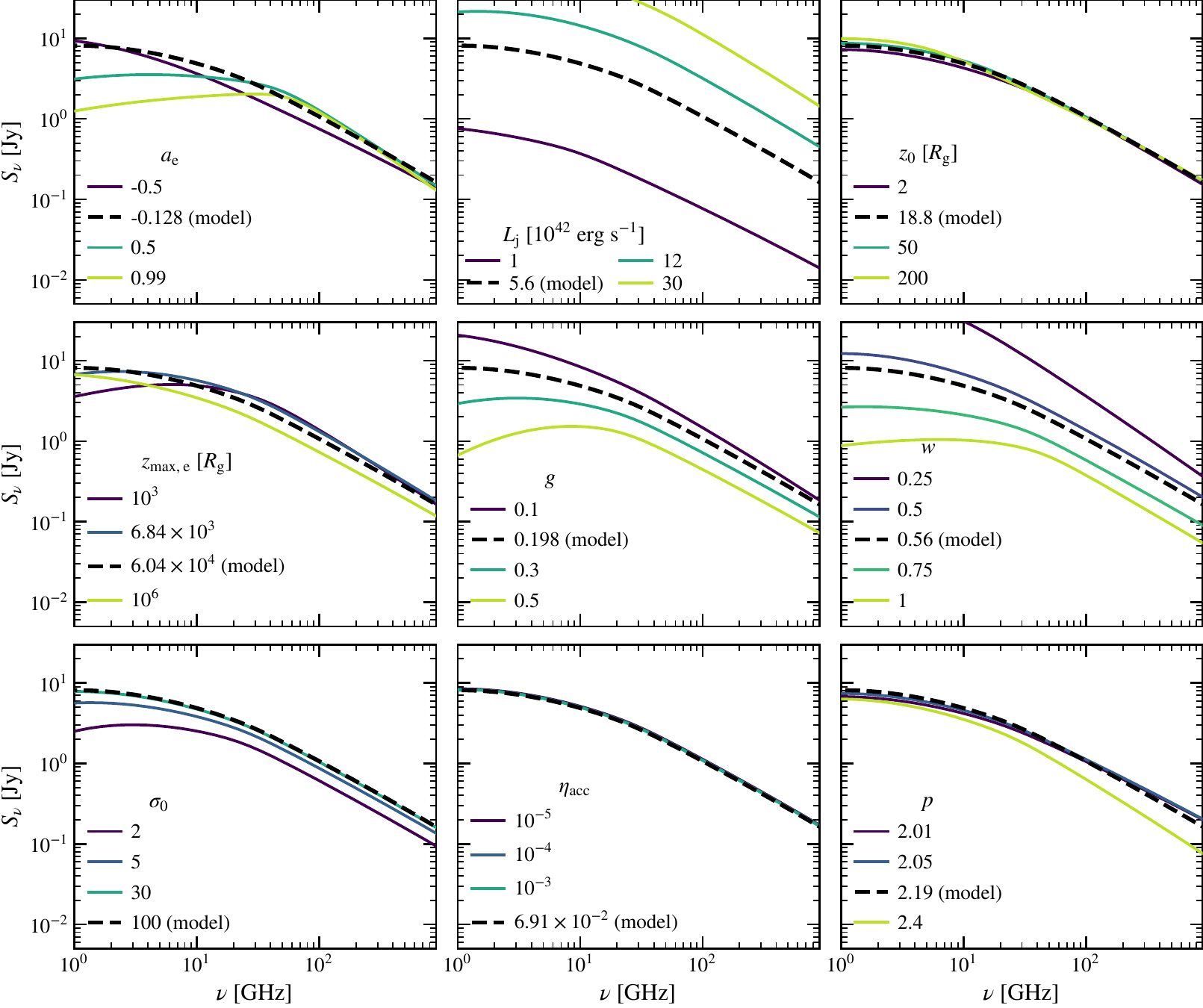}
    \caption{Same as Fig.~\ref{fig:scans_sed}, but for the radio flux density $S_\nu$ between $1$ and $900$~GHz.}
    \label{fig:scans_radio}
\end{figure*}

We show how each parameter of the jet model shapes the SED. For this, we vary one parameter at a time, using the best-fitting jet parameters of the Jet+RIAF model as a reference.
Figs.~\ref{fig:scans_sed} and \ref{fig:scans_radio} show the resulting broadband SEDs and radio spectra of the jet plus counter-jet, integrated over the whole jet and without the RIAF.
For simplicity, we do not show the spectra integrated within the effective apertures that are compared with the data in Sect.~\ref{sec:results}.
The whole-jet radio spectrum of the reference model peaks at ${\approx}1$~GHz and is optically thin above a few tens of GHz, while its optical-to-X-ray emission is optically thin synchrotron emission from cooled electrons.

\textit{Injection profile index $a_{\rm e}$.} Larger values of $a_{\rm e}$ concentrate the injection near $z_0$, where the jet is narrow, the magnetic field is strong and the low-frequency emission is self-absorbed. Increasing $a_{\rm e}$ therefore lowers the low-frequency radio flux and moves the peak of the radio spectrum to higher frequencies, while the millimetre emission changes much less and the optical-to-X-ray emission decreases, because less power is injected into the faster, more strongly Doppler-boosted outer jet.

\textit{Jet power $L_{\rm j}$.} Since both the magnetic energy density and the power injected into electrons scale with $L_{\rm j}$, the synchrotron emission grows faster than linearly with it at all frequencies, and the higher opacity moves the peak of the radio spectrum to higher frequencies. $L_{\rm j}$ therefore sets the overall normalisation of the model.

\textit{Jet base height $z_0$.} For $a_{\rm e}\simeq0$, the same power is distributed evenly in $\log z$ between $z_0$ and $\sim z_{\rm max,e}$, so raising $z_0$ increases the power injected per logarithmic interval of height. In the reference model, with $a_{\rm e}\approx-0.13$ and $z_{\rm max,e}\approx6\times10^{4}\,R_{\rm g}$, most of the power is injected far above the base, and $z_0$ changes the spectrum by less than ${\approx}20$ per cent between 2 and $200\,R_{\rm g}$.

\textit{Maximum injection height $z_{\rm max,e}$.} Extending the injection to larger heights, where the jet is wider and the magnetic field weaker, adds optically thin emission at low frequencies and moves the peak of the radio spectrum to lower frequencies (from ${\approx}6$~GHz for $z_{\rm max,e}=10^{3}\,R_{\rm g}$ to ${\approx}1$~GHz for the reference model), at the expense of the emission at $\gtrsim20$~GHz. Since the total power injected into electrons is fixed, the optical-to-X-ray luminosity increases with $z_{\rm max,e}$ up to the reference value, mainly because the electrons injected further out move with a larger $\Gamma_{\rm j}$ and their emission is more strongly Doppler boosted towards the observer; for $z_{\rm max,e}=10^{6}\,R_{\rm g}$, the injection extends into the wide outer jet, where the field is weak, and the luminosity decreases at all frequencies.

\textit{Bulk acceleration index $g$.} A faster acceleration lowers the comoving magnetic field ($B'\propto\Gamma_{\rm j}^{-1}$) and, for $\Gamma_{\rm j}\gtrsim1/\sin i$, de-boosts the emission of the outer jet. Increasing $g$ therefore suppresses the low-frequency radio emission and moves the radio peak to higher frequencies, whereas the optical-to-X-ray emission, produced closer to the base, changes less: it increases up to $g\approx0.2$ and decreases for faster accelerations.

\textit{Jet geometry index $w$.} The jet width is kept fixed at $z=140\,R_{\rm g}$, so larger $w$ implies a wider jet and a weaker magnetic field ($B'\propto R_{\rm j}^{-1}$) further out. The radio emission is most sensitive to $w$ and decreases sharply as $w$ increases, whereas the optical and X-ray emission change much less.

\textit{Initial magnetisation $\sigma'_0$.} For $\sigma'_0\gtrsim30$, the spectrum is practically independent of $\sigma'_0$, because $B'^2\propto\sigma'/(1+\sigma')$ saturates. For low values, the jet runs out of magnetic energy in its outer part, where the prescribed $\Gamma_{\rm j}(z)$ exceeds $\mu$, and the magnetic field vanishes, which lowers the low-frequency radio emission; the optical-to-X-ray emission, produced closer to the base, decreases less (by 20--30 per cent for $\sigma'_0=2$), since the field is weaker along the whole jet.

\textit{Acceleration efficiency $\eta_{\rm acc}$.} The acceleration efficiency sets the maximum energy of the electrons through the balance between acceleration and cooling, and therefore the high-energy cutoff of the synchrotron spectrum: lowering $\eta_{\rm acc}$ moves the cutoff from above $10^{20}$~Hz into the X-ray band and reduces the X-ray luminosity. Since the injected power is fixed, the power removed at the highest energies is transferred to lower-energy electrons, slightly increasing the radio flux.

\textit{Electron spectral index $p$.} Steeper injection spectra put a larger fraction of the injected power into low-energy electrons, so the optical-to-X-ray spectrum steepens and its luminosity decreases strongly, especially in X-rays, while the radio emission changes much less.

In summary, the radio spectrum mainly depends on the parameters that control the geometry, magnetic field and bulk motion of the outer jet ($a_{\rm e}$, $z_{\rm max,e}$, $g$, $w$), whereas the optical and X-ray emission mainly depends on $L_{\rm j}$ (together with $\epsilon_{\rm e}$), $p$ and $\eta_{\rm acc}$. The radio data at different angular scales, which sample the jet at increasing heights, are therefore essential to constrain the first group.

\section{Fitting the jet parameters}
\label{app:fits}

In each scenario, we sample the posterior distribution of the eight jet parameters of Sect.~\ref{subsec:obs_constraints} with \textsc{emcee} \citep{ForemanMackey_etal2013} through \textsc{bilby} \citep{Ashton_etal2019}, with 48 walkers and 3000 steps. The priors are uniform within the ranges of Table~\ref{tab:parameters} (uniform in the logarithm for $L_{\rm j}$, $\epsilon_{\rm e}$, $z_0$, $z_{\rm max,e}$ and $\eta_{\rm acc}$). The other parameters are fixed as in Table~\ref{tab:parameters}, and the gravitational redshift and the effective apertures of Sect.~\ref{subsec:obs_constraints} are included. The walkers start near the best model of an earlier, wider exploration (see below), and we adopt a burn-in fraction of $0.3$. Each fit computed ${\approx}9\times10^{4}$ models. The chains are stationary: the acceptance fraction is 0.30--0.32; the burn-in exceeds three times the integrated autocorrelation time $\tau$ of every parameter ($\tau\le128$ and 151 steps in the Jet-only and Jet+RIAF chains); the chains are 23 and 20 times $\tau$ long, which gives ${\approx}780$ and 670 independent samples; and the medians of the first and last thirds of the retained chain differ by less than 0.35 standard deviations for every parameter, and by less than 0.2 for those that the data constrain.

\begin{figure*}
    \centering
    \includegraphics[width=0.85\textwidth]{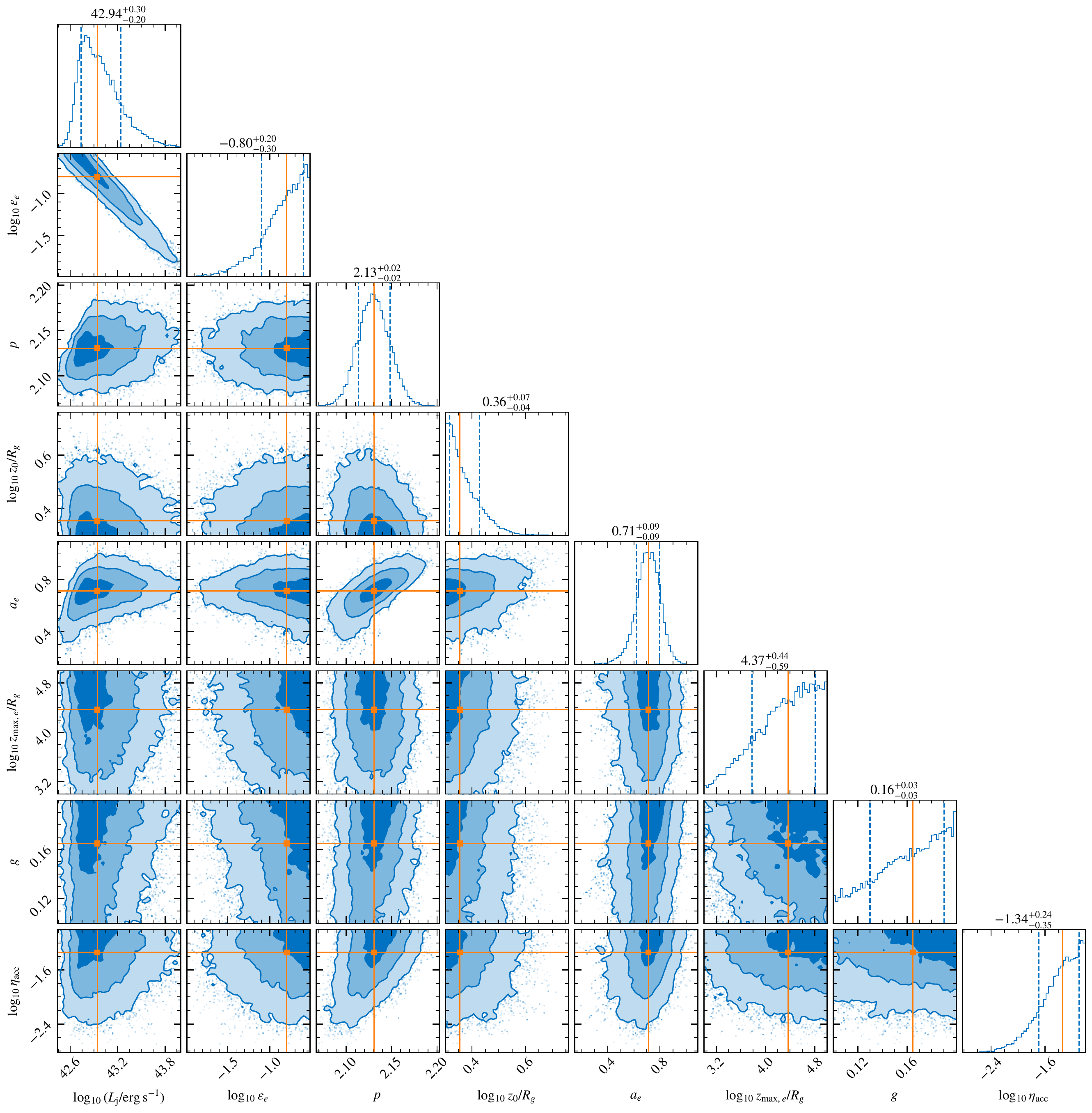}
    \caption{Posterior distribution of the jet parameters of the Jet-only fit (Appendix~\ref{app:fits}), after the burn-in, in the sampled variables (the decimal logarithm for the parameters with logarithmic priors in Table~\ref{tab:parameters}).
    The orange lines mark the posterior medians, and the title of each column gives the median with the 16th and 84th percentiles (dashed lines in the one-dimensional histograms). The contours enclose 39, 86 and 99 per cent of the samples.}
    \label{fig:app_corner_jo}
\end{figure*}

\begin{figure*}
    \centering
    \includegraphics[width=0.85\textwidth]{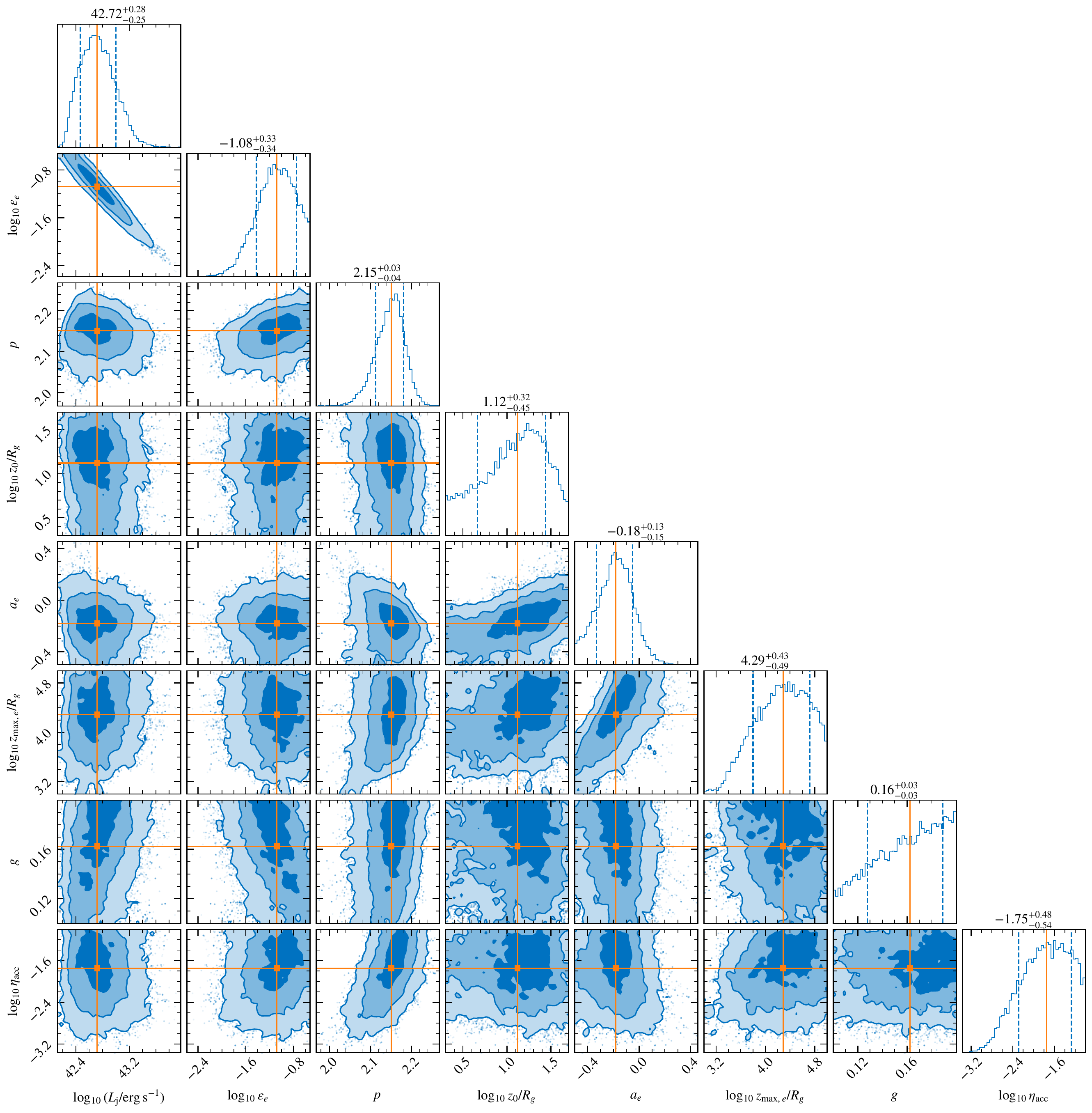}
    \caption{Same as Fig.~\ref{fig:app_corner_jo}, for the Jet+RIAF fit.}
    \label{fig:app_corner_jr}
    \label{lastpage}%
\end{figure*}

Table~\ref{tab:parameters} lists the posterior medians and 68 per cent intervals, and Figs.~\ref{fig:app_corner_jo} and \ref{fig:app_corner_jr} show the posterior distributions.
The best-fitting sample need not lie close to the medians of the marginal distributions: in the Jet+RIAF fit, its $p$, $z_{\rm max,e}$, $g$ and $\eta_{\rm acc}$ are above their 84th percentiles.
Both scenarios require $p\approx2.13$--$2.15$, set by the slope of the X-ray spectrum, and an acceleration efficiency of a few per cent ($\eta_{\rm acc}=0.046^{+0.034}_{-0.025}$ and $0.018^{+0.036}_{-0.013}$), which places the synchrotron cutoff above the \textit{NuSTAR} band. $L_{\rm j}$ and $\epsilon_{\rm e}$ are strongly anticorrelated (with correlation coefficients of $-0.94$ and $-0.97$ between their logarithms), since the synchrotron emission depends mainly on their product, but the fit still constrains $L_{\rm j}$ to within a factor of ${\approx}2$ (Sect.~\ref{subsec:phys_cond}). The posteriors of $g$ and $z_{\rm max,e}$ are close to their priors. The two scenarios differ in the injection profile, with $a_{\rm e}=0.71\pm0.09$ in the Jet-only and $-0.18^{+0.13}_{-0.15}$ in the Jet+RIAF fit, and above all in the height of the base of the jet.

In the Jet-only fit, the posterior of $z_0$ piles up against the hard lower limit of the prior, $2\,R_{\rm g}$ (Fig.~\ref{fig:app_corner_jo}), and $z_0<3.6\,R_{\rm g}$ at the 99 per cent level.
The fit worsens quickly if the base is forced higher: the best models computed with $z_0=6$--$8$ and 10--$13\,R_{\rm g}$ have $\chi^2_{\rm red}=1.9$ and 2.9. In the Jet+RIAF fit, the posterior of $z_0$ is close to the prior (Fig.~\ref{fig:app_corner_jr}): the RIAF makes the $230$~GHz emission independent of the base of the jet.

Over the Jet-only posterior, the jet provides $481^{+72}_{-67}$~mJy within the EHT aperture (a model-to-data ratio of $0.72^{+0.11}_{-0.10}$), of which 81 per cent is emitted below $z=10\,R_{\rm g}$ and 35 per cent by the counter-jet; the flux-weighted gravitational redshift factor is $g_{\rm r}\approx0.85$. The Jet-only model also underestimates the arcsecond-scale $230$~GHz flux (an ALMA model-to-data ratio of $0.79\pm0.07$). The Jet+RIAF model reproduces both scales (1.06 and $1.02^{+0.09}_{-0.07}$); in it, the jet contributes $1.4^{+1.3}_{-0.6}$ per cent of the $230$~GHz flux within the EHT aperture, 45 per cent of the whole-source $230$~GHz flux, 80 per cent of the optical flux at $5.2\times10^{14}$~Hz and 95 per cent of the X-ray flux at 0.45~keV.

These results are robust: we reach the same conclusions under less restrictive fitting conditions. An alternative MCMC, in which $z_{\rm acc}$ (50--$1000\,R_{\rm g}$) was also free, $g$ had a wider range (0.1--0.5) and the jet emission was integrated up to $z_*=R_*/\sin i$ in each aperture, gives the same posteriors for the common parameters. A differential-evolution search with the width index $w$ also free lowers the best Jet-only $\chi^2_{\rm red}$ to 1.05, but only for a jet 2.6--3.6 times wider than observed at $10^4$--$10^5\,R_{\rm g}$ \citep{Nakamura_etal2018}.

\end{document}